\documentclass[aps,twocolumn,showpacs]{revtex4}
\usepackage{epsf}
\usepackage{bm}

\begin{document}

\title{Strong exciton-plasmon coupling in semiconducting carbon nanotubes}

\author{I.V.~Bondarev}\email[Corresponding author.
E-mail: ]{ibondarev@nccu.edu}\affiliation{Physics Department,
North Carolina Central University, 1801 Fayetteville Str, Durham,
NC 27707, USA}
\author{L.M.~Woods}
\author{K.~Tatur}\affiliation{Physics Department, University of South Florida,
4202 E.Fowler Ave, Tampa, FL 33620, USA}

\begin{abstract}
We study theoretically the interactions of excitonic states with
surface electromagnetic modes of small-diameter
($\,\lesssim\!1$~nm) semiconducting single-walled carbon
nanotubes. We show that these interactions can result in strong
exciton-surface-plasmon coupling. The exciton absorption line
shape exhibits Rabi splitting $\sim0.1$~eV as the exciton energy
is tuned to the nearest interband surface plasmon resonance of the
nanotube. We also show that the quantum confined Stark effect may
be used as a tool to control the exciton binding energy and the
nanotube band gap in carbon nanotubes in order, e.~g., to bring
the exciton total energy in resonance with the nearest interband
plasmon mode.~The exciton-plasmon Rabi splitting we predict here
for an individual carbon nanotube is close in its magnitude to
that previously reported for hybrid plasmonic nanostructures
artificially fabricated of organic semiconductors on metallic
films. We expect this effect to open up paths to new tunable
optoelectronic device applications of semiconducting carbon
nanotubes.
\end{abstract}

\maketitle

\section{Introduction}

Single-walled carbon nanotubes (CNs) are quasi-one-dimensional
(1D) cylindrical wires consisting of graphene sheets rolled-up
into cylinders with diameters $\sim\!1-10$~nm and lengths
$\sim\!1-10^4\,~\mu$m~\cite{Dresselhaus,Dai,Zheng,Huang}. CNs are
shown~to~be useful as miniaturized electronic, electromechanical,
and chemical devices~\cite{Baughman}, scanning probe
devices~\cite{Popescu}, and nanomaterials for macroscopic
composites~\cite{Trancik}. The area of their potential
applications was recently expanded to
nanophotonics~\cite{Bondarev06,Bondarev07,Bondarev07jem,Bondarev07os}
after the demonstration of controllable single-atom incapsulation
into CNs~\cite{Shimoda,Jeong,Jeongtsf,Khazaei}, and even to
quantum cryptography since the experimental evidence was reported
for quantum correlations in the photoluminescence spectra of
individual nanotubes~\cite{Imamoglu}.

For pristine (undoped) single-walled CNs, the numerical
calculations predicting large exciton binding energies
($\sim\!0.3\!-\!0.6$~eV) in semiconducting
CNs~\cite{Pedersen03,Pedersen04,Capaz}~and even in some
small-diameter ($\sim\!0.5$~nm) metallic CNs~\cite{Spataru04},
followed by the results of various exciton photoluminescence
measurements~\cite{Imamoglu,Wang04,Wang05,Hagen05,Plentz05,Avouris08},
have become available.~These works, together with other reports
investigating the role of effects such as intrinsic
defects~\cite{Hagen05,Prezhdo08}, exciton-phonon
interactions~\cite{Plentz05,Prezhdo08,Perebeinos05,Lazzeri05,Piscanec},
external magnetic and electric
fields~\cite{Zaric,Perebeinos07,Srivastava}, reveal the variety
and complexity of the intrinsic optical properties of
CNs~\cite{Dresselhaus07}.

Here we develop a theory for the interactions between excitonic
states and surface electromagnetic (EM) modes in small-diameter
($\lesssim\!1$~nm) semiconducting single-walled CNs.~We
demonstrate that such interactions can result in a strong
exciton-surface-plasmon coupling due to the presence of low-energy
($\sim\!0.5\!-\!2$~eV) weakly-dispersive inter\-band plasmon
modes~\cite{Pichler98} and large exciton excitation energies
$\sim\!1$~eV~\cite{Spataru05,Ma} in small-diameter CNs. Previous
studies have been focused on artificially fabricated \emph{hybrid}
plasmonic nanostructures, such as dye molecules in organic
polymers deposited on metallic films~\cite{Bellessa},
semiconductor quantum dots coupled to metallic
nanoparticles~\cite{Govorov}, or nanowires~\cite{Fedutik}, where
one material carries the exciton and another one carries the
plasmon. Our results are particularly interesting since they
reveal the fundamental EM phenomenon --- the strong
exciton-plasmon coupling --- in an \emph{individual} quasi-1D
nanostructure, a~carbon nanotube.

The paper is organized as follows.~Section~\ref{sec2} presents the
general Hamiltonian of the exciton interaction with vacuum-type
quantized surface EM modes of a single-walled CN. No external EM
field is assumed to be applied.~The vacuum--type--field we
consider is created by CN surface EM fluctuations. In describing
the exciton--field interaction on the CN surface, we use our
recently developed Green function formalism to quantize the EM
field in the presence of quasi-1D absorbing
bodies~\cite{Bondarev02,Bondarev04,Bondarev04pla,Bondarev04ssc,Bondarev05,Bondarev06trends}.
The formalism follows the original line of the macroscopic quantum
electrodynamics (QED) approach developed by Welsch and coworkers
to correctly describe medium-assisted electromagnetic vacuum
effects in dispersing and absorbing
media~\cite{VogelWelsch,ScheelWelsch,BuhmannWelsch} (also refs.
therein).~Section~\ref{sec3} explains how the interaction
introduced in Sec.~II results in the coupling of the excitonic
states to the nanotube's surface plasmon modes.~Here we derive,
calculate and discuss the characteristics of the coupled
exciton--plasmon excitations, such as the dispersion relation, the
plasmon density of states (DOS), and the optical absorption line
shape, for particular semiconducting CNs of different diameters.
We also analyze how the electrostatic field applied perpendicular
to the CN axis affects the CN band gap, the exciton binding
energy, and the surface plasmon energy, to explore the tunability
of the exciton-surface-plasmon coupling in CNs.~The summary and
conclusions of the work are given in Sec.~\ref{concl}. All the
technical details about the construction and diagonalization of
the exciton--field Hamiltonian, the EM field Green tensor
derivation, the perpendicular electrostatic field effect, are
presented in the Appendices in order not to interrupt the flow of
the arguments and results.

\section{Exciton-electromagnetic-field interaction on the
nanotube surface}\label{sec2}

\begin{figure}[t]
\epsfxsize=8.65cm\centering{\epsfbox{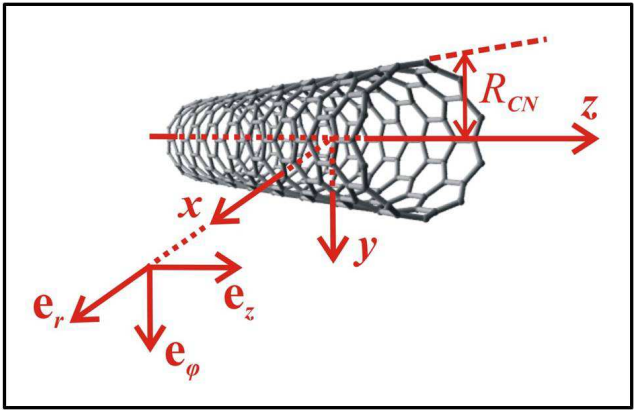}}\caption{(Color
online) The geometry of the problem.}\label{fig1}
\end{figure}

We consider the vacuum-type EM interaction of an exciton with the
quantized surface electromagnetic fluctuations of a single-walled
semiconducting CN by using our recently developed Green function
formalism to quantize the EM field in the presence of quasi-1D
absorbing
bodies~\cite{Bondarev02,Bondarev04,Bondarev04pla,Bondarev04ssc,Bondarev05,Bondarev06trends}.
No external EM field is assumed to be applied. The nanotube is
modelled by an infinitely thin, infinitely long, anisotropically
conducting cylinder with its surface conductivity obtained from
the realistic band structure of a particular CN. Since the problem
has the cylindrical symmetry, the orthonormal cylindrical basis
$\{\mathbf{e}_{r},\mathbf{e}_{\varphi},\mathbf{e}_{z}\}$ is used
with the vector $\mathbf{e}_{z}$ directed along the nanotube axis
as shown in Fig.~\ref{fig1}. Only the axial conductivity,
$\sigma_{zz}$, is taken into account, whereas the azimuthal one,
$\sigma_{\varphi\varphi}$, being strongly suppressed by the
transverse depolarization
effect~\cite{Benedict,Tasaki,Li,Marinop,Ando,Kozinsky}, is
neglected.

The total Hamiltonian of the coupled exciton-photon system on the
nanotube surface is of the form
\begin{equation}
\hat{H}=\hat{H}_F+\hat{H}_{ex}+\hat{H}_{int},\label{Htot}
\end{equation}
where the three terms represent the free (medium-assisted) EM
field, the free (non-interacting) exciton, and their interaction,
respectively.~More explicitly, the second quantized field
Hamiltonian is
\begin{equation}
\hat{H}_F\!=\!\sum_\mathbf{n}\int_0^\infty\!\!\!d\omega\,\hbar\omega
\hat{f}^\dag(\mathbf{n},\omega)\hat{f}(\mathbf{n},\omega),
\label{HF}
\end{equation}
where the scalar bosonic field operators
$\hat{f}^\dag(\mathbf{n},\omega)$ and $\hat{f}(\mathbf{n},\omega)$
create and annihilate, respectively, the surface EM excitation of
frequency $\omega$ at an arbitrary point
$\mathbf{n}\!=\!\mathbf{R}_n\!=\!\{R_{CN},\varphi_n,z_n\}$
associated with a carbon atom (representing a lattice site --
Fig.~\ref{fig1}) on the surface of the CN of radius $R_{CN}$. The
summation is made over all the carbon atoms, and in the following
it is replaced by the integration over the entire nanotube surface
according to the rule
\begin{equation}
\sum_\mathbf{n}\!\ldots=\frac{1}{S_0}\int\!d\mathbf{R}_n\!\ldots=
\frac{1}{S_0}\int_0^{2\pi}\!\!\!\!d\varphi_nR_{CN}\!\int_{-\infty}^\infty\!\!\!\!dz_n\!\ldots,
\label{sumrule}
\end{equation}
where $S_0\!=\!(3\sqrt{3}/4)b^2$ is the area of an elementary
equilateral triangle selected around each carbon atom in a way to
cover the entire surface of the nanotube, $b\!=\!1.42$~\AA\space
is the carbon-carbon interatomic distance.

\begin{figure}[t]
\epsfxsize=8.65cm\centering{\epsfbox{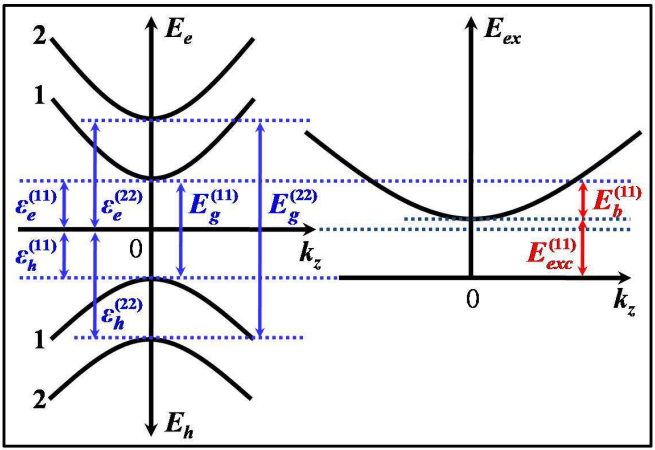}}\caption{(Color
online)~Schematic of the two transversely quantized azimuthal
electron-hole subbands (\emph{left}), and the first-interband
ground-internal-state exciton energy (\emph{right}) in a
small-diameter semiconducting carbon nanotube. Subbands with
indices $j=1$ and 2 are shown, along with the optically allowed
(exciton-related) interband transitions~\cite{Ando}. See text for
notations.}\label{fig2}
\end{figure}

The second quantized Hamiltonian of the free exciton (see, e.g.,
Ref.~\cite{Haken}) on the CN surface is of the form
\begin{equation}
\hat{H}_{ex}\!=\!\!\!\sum_{\mathbf{n},\mathbf{m},f}\!\!\!E_f(\mathbf{n})
B^\dag_{\mathbf{n}+\mathbf{m},f}B_{\mathbf{m},f}\!=
\!\sum_{\mathbf{k},f}E_f(\mathbf{k})B^\dag_{\mathbf{k},f}B_{\mathbf{k},f},
\label{Hex}
\end{equation}
where the operators $B^\dag_{\mathbf{n},f}$ and $B_{\mathbf{n},f}$
create and annihilate, respectively, an exciton with the energy
$E_f(\mathbf{n})$ in the lattice site $\mathbf{n}$ of the CN
surface.~The index $f\,(\ne\!0)$ refers to the internal degrees of
freedom of the exciton. Alternatively,
\begin{equation}
B^\dag_{\mathbf{k},f}=\frac{1}{\sqrt{N}}\sum_{\mathbf{n}}\!B^\dag_{\mathbf{n},f}
e^{i\mathbf{k}\cdot\mathbf{n}}~~~\mbox{and}~~~B_{\mathbf{k},f}=(B^\dag_{\mathbf{k},f})^\dag
\label{Bkf}
\end{equation}
create and annihilate the $f$-internal-state exciton with the
quasi-momentum $\mathbf{k}\!=\!\{k_{\varphi},k_{z}\}$, where the
azimuthal component is quantized due to the transverse confinement
effect and the longitudinal one is continuous, $N$ is the total
number of the lattice sites (carbon atoms) on the CN surface. The
exciton total energy is then written in the form
\begin{equation}
E_f(\mathbf{k})=E_{exc}^{(f)}(k_{\varphi})+\frac{\hbar^2k_z^2}{2M_{ex}(k_{\varphi})}
\label{Ef}
\end{equation}
Here, the first term represents the excitation energy
\begin{equation}
E_{exc}^{(f)}(k_{\varphi})=E_g(k_{\varphi})+E_b^{(f)}(k_{\varphi})
\label{Eexcf}
\end{equation}
of the $f$-internal-state exciton with the (negative) binding
energy $E_b^{(f)}$, created via the interband transition with the
band gap
\begin{equation}
E_g(k_{\varphi})=\varepsilon_e(k_{\varphi})+\varepsilon_h(k_{\varphi}),
\label{Egkfi}
\end{equation}
where $\varepsilon_{e,h}$ are transversely quantized azimuthal
electron-hole subbands (see the schematic in Fig.~\ref{fig2}).~The
second term in Eq.~(\ref{Ef}) represents the kinetic energy of the
translational longitudinal movement of the exciton with the
effective mass $M_{ex}=m_e+m_h$, where $m_e$ and $m_h$ are the
(subband-dependent) electron and hole effective masses,
respectively. The two equivalent free-exciton Hamiltonian
representations are related to one another via the obvious
orthogonality relationships
\begin{equation}
\frac{1}{N}\sum_\mathbf{n}e^{-i(\mathbf{k}-\mathbf{k}^\prime)\cdot\mathbf{n}}=
\delta_{\mathbf{k}\mathbf{k}^\prime},\;\;
\frac{1}{N}\sum_\mathbf{k}e^{-i(\mathbf{n}-\mathbf{m})\cdot\mathbf{\mathbf{k}}}=
\delta_{\mathbf{n}\mathbf{m}} \label{orthog}
\end{equation}
with the $\mathbf{k}$-summation running over the first Brillouin
zone of the nanotube.~The bosonic field operators in $\hat{H}_F$
are transformed to the $\mathbf{k}$-representation in the same
way.

The most general (non-relativistic, electric dipole)
exciton-photon interaction on the nanotube surface can be written
in the form (we use the Gaussian system of units and the Coulomb
gauge; see details in Appendix~\ref{appA})
\begin{eqnarray}
\hat{H}_{int}=\sum_{\mathbf{n},\mathbf{m},f}\int_{0}^{\infty}\!\!\!\!\!d\omega\,[\,
\mbox{g}_f^{(+)}(\mathbf{n},\mathbf{m},\omega)B^\dag_{\mathbf{n},f}\hskip0.6cm\nonumber\\[0.1cm]
-\;\mbox{g}_f^{(-)}(\mathbf{n},\mathbf{m},\omega)B_{\mathbf{n},f}\,]\,
\hat{f}(\mathbf{m},\omega)+h.c.,\label{Hint}
\end{eqnarray}
where
\begin{equation}
\mbox{g}_f^{(\pm)}(\mathbf{n},\mathbf{m},\omega)=
\mbox{g}_f^{\perp}(\mathbf{n},\mathbf{m},\omega)\pm
{\omega\over{\omega_f}}\,\mbox{g}_f^{\parallel}(\mathbf{n},\mathbf{m},\omega)
\label{gpm}
\end{equation}
with
\begin{eqnarray}
\mbox{g}_f^{\perp(\parallel)}(\mathbf{n},\mathbf{m},\omega)=
-i\frac{4\omega_f}{c^{2}}\sqrt{\pi\hbar\omega\,\mbox{Re}\,\sigma_{zz}(R_{CN},\omega)}
\hskip0.5cm\nonumber\\[0.2cm]
\times\;\;(\textbf{d}^f_{\mathbf{n}})_z\,^{\perp(\parallel)}G_{zz}(\mathbf{n},\mathbf{m},\omega)
\hskip1.4cm\label{gperppar}
\end{eqnarray}
being the interaction matrix element where the exciton with the
energy $E_{exc}^{(f)}=\hbar\omega_f$ is excited through the
electric dipole transition
$(\textbf{d}^f_{\mathbf{n}})_z\!=\!\langle0|(\hat{\mathbf{d}}_\mathbf{n})_z|f\rangle$
in the lattice site $\textbf{n}$ by the nanotube's transversely
(longitudinally) polarized surface EM modes. The modes are
represented in the matrix element by the transverse (longitudinal)
part of the Green tensor $zz$-component
$G_{zz}(\mathbf{n},\mathbf{m},\omega)$ of the EM subsystem
(Appendix~B). This is the only Green tensor component we have to
take into account.~All the other components can be safely
neglected as they are greatly suppressed by the strong transverse
depolarization effect in
CNs~\cite{Benedict,Tasaki,Li,Marinop,Ando,Kozinsky}.~As a
consequence, only $\sigma_{zz}(R_{CN},\omega)$, the \emph{axial}
dynamic surface conductivity per unit length, is present in
Eq.(\ref{gperppar}).

Equations~(\ref{Htot})--(\ref{gperppar}) form the complete set of
equations describing the exciton-photon coupled system on the CN
surface in terms of the EM field Green tensor and the CN surface
axial conductivity.

\section{Exciton-surface-plasmon coupling}\label{sec3}

For the following it is important to realize that the transversely
polarized surface EM mode contribution to the
interaction~(\ref{Hint})--(\ref{gperppar}) is negligible compared
to the longitudinally polarized surface EM mode contribution. As
a~matter of fact,
$^{\perp}G_{zz}(\mathbf{n},\mathbf{m},\omega)\!\equiv\!0$ in the
model of an infinitely thin cylinder we use here
(Appendix~\ref{appB}), thus yielding
\begin{equation}
\mbox{g}_f^\perp(\mathbf{n},\mathbf{m},\omega)\!\equiv\!0,~~~
\mbox{g}_f^{(\pm)}(\mathbf{n},\mathbf{m},\omega)\!=\!
\pm\frac{\omega}{\omega_f}\,\mbox{g}_f^{\parallel}(\mathbf{n},\mathbf{m},\omega)
\label{gfpar}
\end{equation}
in Eqs.~(\ref{Hint})--(\ref{gperppar}).~The point is that, because
of the nanotube quasi-one-dimensionality, the exciton
quasi-momentum vector and all the relevant vectorial matrix
elements of the momentum and dipole moment operators are directed
predominantly along the CN axis (the longitudinal exciton; see,
however, Ref.~\cite{UryuAndo}). This prevents the exciton from the
electric dipole coupling to transversely polarized surface EM
modes as they propagate predominantly along the CN axis with their
electric vectors orthogonal to the propagation direction.~The
longitudinally polarized surface EM modes are generated by the
electronic Coulomb potential (see, e.g., Ref.~\cite{Landau}), and
therefore represent the CN surface plasmon excitations.~These have
their electric vectors directed along the propagation direction.
They do couple to the longitudinal excitons on the CN surface.
Such modes were observed in Ref.~\cite{Pichler98}. They occur in
CNs both at high energies (well-known $\pi$-plasmon~at
$\sim\!6$~eV) and at comparatively low energies of
$\sim\!0.5\!-\!2$~eV.~The latter ones are related to the
transversely quantized interband (inter-van Hove) electronic
transitions. These weakly-dispersive
modes~\cite{Pichler98,Kempa02} are similar to the intersubband
plasmons in quantum wells~\cite{Kempa89}. They occur in the same
energy range of $\sim\!1$~eV where the exciton excitation energies
are located in small-diameter ($\lesssim\!1$~nm) semiconducting
CNs~\cite{Spataru05,Ma}. In what follows we focus our
consideration on the exciton interactions with these particular
surface plasmon modes.

\subsection{The dispersion relation}

To obtain the dispersion relation of the coupled exci\-ton-plasmon
excitations, we transfer the total
Hamiltonian~(\ref{Htot})--(\ref{Hint}) and (\ref{gfpar}) to the
$\mathbf{k}$-representation using Eqs.~(\ref{Bkf}) and
(\ref{orthog}), and then diagonalize it exactly by means of
Bogoliubov's canonical transformation technique (see, e.g.,
Ref.~\cite{Davydov}). The details of the procedure are given in
Appendix~\ref{appC}. The Hamiltonian takes the form
\begin{equation}
\hat{H}=\!\!\!\sum_{\mathbf{k},\,\mu=1,2}\!\!\!\hbar\omega_\mu(\mathbf{k})\,
\hat{\xi}^\dag_\mu(\mathbf{k})\hat{\xi}_\mu(\mathbf{k})+E_0\,.
\label{Htotdiag}
\end{equation}
Here, the new operator
\begin{eqnarray}
\hat{\xi}_\mu(\mathbf{k})=
\sum_{f}\left[u_\mu^\ast(\mathbf{k},\omega_f)B_{\mathbf{k},f}
-v_\mu(\mathbf{k},\omega_f)B_{-\mathbf{k},f}^\dag\right]\label{xik}\\
+\int_0^\infty\!\!\!\!\!d\omega\left[u_\mu(\mathbf{k},\omega)\hat{f}(\mathbf{k},\omega)-
v_\mu^\ast(\mathbf{k},\omega)\hat{f}^\dag(-\mathbf{k},\omega)\right]\nonumber
\end{eqnarray}
annihilates and
$\hat{\xi}^\dag_\mu(\mathbf{k})\!=\![\hat{\xi}_\mu(\mathbf{k})]^\dag$
creates the exciton-plas\-mon excitation of branch $\mu$, the
quantities $u_\mu$ and $v_\mu$ are appropriately chosen canonical
transformation coefficients. The "vacuum" energy $E_0$ represents
the state with no exciton-plasmons excited in the system, and
$\hbar\omega_\mu(\mathbf{k})$ is the exciton-plasmon energy given
by the solution of the following (dimensionless) dispersion
relation
\begin{equation}
x_\mu^2-\varepsilon_f^2-\varepsilon_f\frac{2}{\pi}\int_0^\infty\!\!\!\!\!dx\,
\frac{x\,\bar\Gamma_0^f(x)\rho(x)}{x_\mu^2-x^2}=0\,.\label{dispeq}
\end{equation}
Here,
\begin{equation}
x=\frac{\hbar\omega}{2\gamma_0},~~~
x_\mu=\frac{\hbar\omega_\mu(\mathbf{k})}{2\gamma_0},~~~
\varepsilon_f=\frac{E_f(\mathbf{k})}{2\gamma_0}
\label{dimless}
\end{equation}
with $\gamma_0\!=\!2.7$~eV being the carbon nearest neighbor
overlap integral entering the CN surface axial conductivity
$\sigma_{zz}(R_{CN},\omega)$. The function
\begin{equation}
\bar\Gamma_0^f(x)=\frac{4|d^f_z|^2x^3}{3\hbar c^3}
\left(\frac{2\gamma_0}{\hbar}\right)^{\!2}
\label{Gamma0f}
\end{equation}
with
$d^f_z\!=\!\sum_{\mathbf{n}}\langle0|(\hat{\mathbf{d}}_\mathbf{n})_z|f\rangle$
represents the (dimensionless) spontaneous decay rate, and
\begin{equation}
\rho(x)=\frac{3S_0}{16\pi\alpha
R_{CN}^2}\;\mbox{Re}\frac{1}{\bar\sigma_{zz}(x)}\label{plDOS}
\end{equation}
stands for the surface plasmon density of states (DOS) which is
responsible for the exciton decay rate variation due to its
coupling to the plasmon modes.~Here, $\alpha\!=\!e^2/\hbar
c\!=\!1/137$ is the fine-structure constant and
$\bar\sigma_{zz}\!=\!2\pi\hbar\sigma_{zz}/e^2$ is the
dimensionless CN surface axial conductivity per unit length.

Note that the conductivity factor in Eq.~(\ref{plDOS}) equals
\begin{equation}
\mbox{Re}\frac{1}{\bar\sigma_{zz}(x)}=-\frac{4\alpha
c}{R_{CN}}\left(\frac{\hbar}{2\gamma_0x}\right)\mbox{Im}\frac{1}{\epsilon_{zz}(x)-1}
\label{Reoneoversigma}
\end{equation}
in view of Eq.~(\ref{dimless}) and equation
$\sigma_{zz}\!=\!-i\omega(\epsilon_{zz}\!-\!1)/4\pi S\rho_{T}$
representing the Drude relation for CNs, where $\epsilon_{zz}$ is
the longitudinal (along the CN axis) dielectric function, $S$ and
$\rho_{T}$ are the surface area of the tubule and the number of
tubules per unit volume,
respectively~\cite{Bondarev04,Bondarev05,Tasaki}.~This relates
very closely the surface plasmon DOS function~(\ref{plDOS}) to the
loss function $-\mbox{Im}(1/\epsilon)$ measured in Electron Energy
Loss Spectroscopy (EELS) experiments to determine the properties
of collective electronic excitations in solids~\cite{Pichler98}.

\begin{figure}[t]
\epsfxsize=8.65cm\centering{\epsfbox{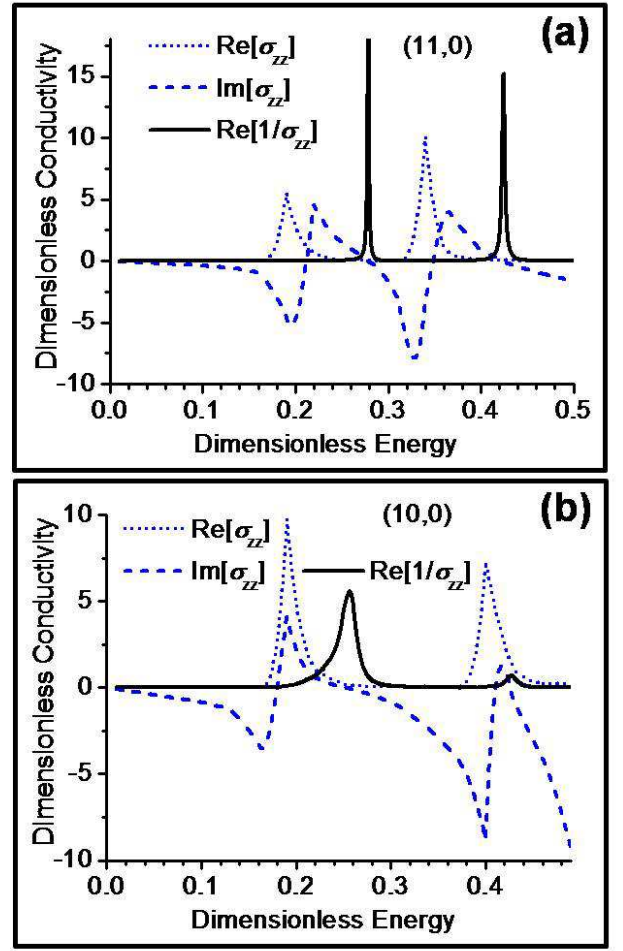}}\caption{(Color
online)~(a),(b)~Calculated dimensionless (see text) axial surface
conductivities for the (11,0) and (10,0) CNs. The dimensionless
energy is defined as [\emph{Energy}]/$2\gamma_0$, according to
Eq.~(\ref{dimless}).}\label{fig3}
\end{figure}

\begin{figure*}[t]
\epsfxsize=17.9cm\centering{\epsfbox{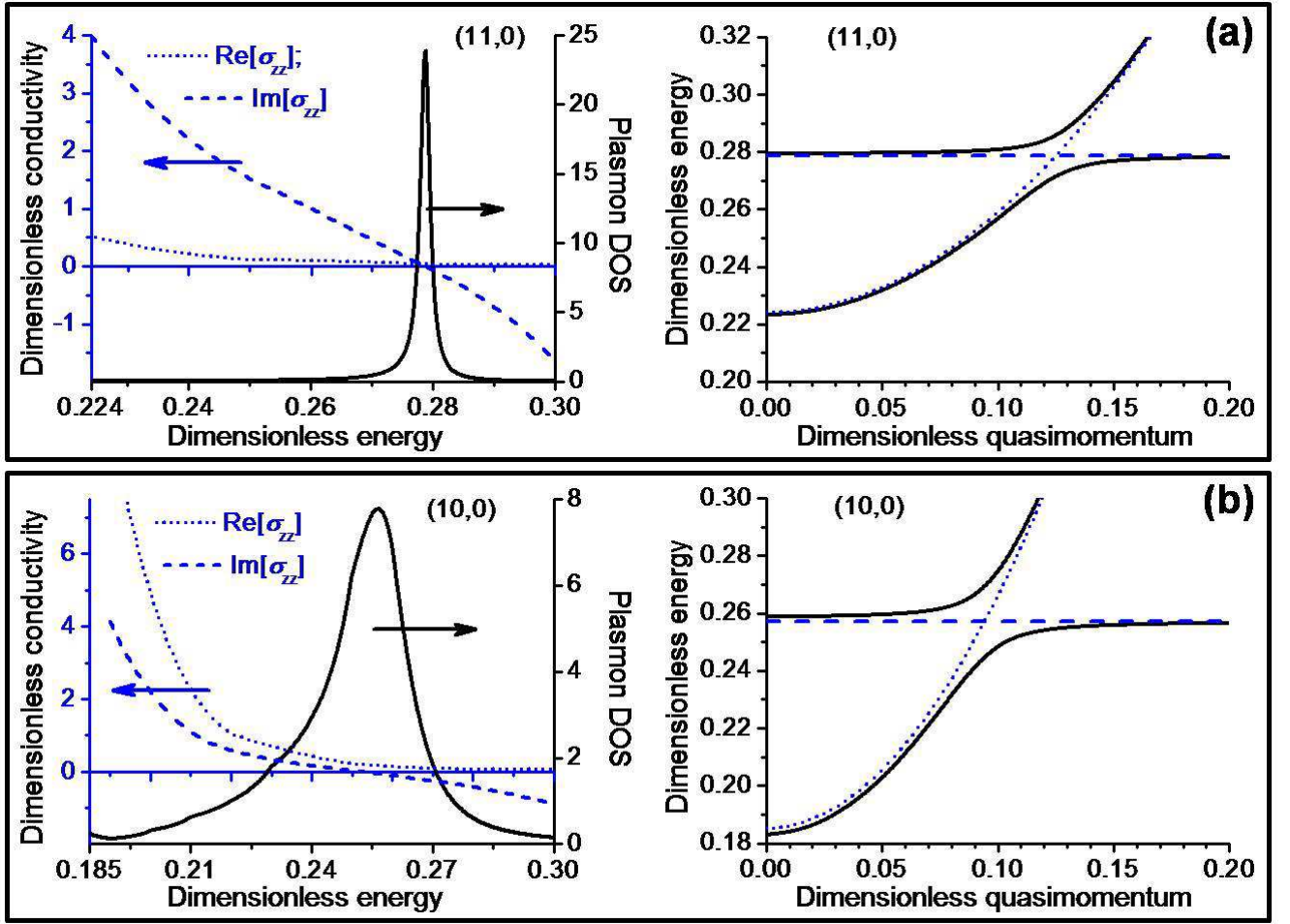}}\caption{(Color
online)~(a),(b)~Surface plasmon DOS and conductivities (left
panels), and lowest bright exciton dispersion when coupled to
plasmons (right panels) in (11,0) and (10,0) CNs, respectively.
The dimensionless energy is defined as
[\emph{Energy}]/$2\gamma_0$, according to Eq.~(\ref{dimless}). See
text for the dimensionless quasi-momentum.}\label{fig4}
\end{figure*}

Figure~\ref{fig3} shows the low-energy behaviors of the functions
$\bar\sigma_{zz}(x)$ and $\mbox{Re}[1/\bar\sigma_{zz}(x)]$ for the
(11,0) and (10,0) CNs ($R_{CN}=0.43$~nm and $0.39$~nm,
respectively) we study here. We obtained them numerically as
follows. First, we adapt the nearest-neighbor non-orthogonal
tight-binding approach~\cite{Valentin} to determine the realistic
band structure of each CN. Then, the room-temperature longitudinal
dielectric functions $\epsilon_{zz}$ are calculated within the
random-phase approximation~\cite{LinShung97,EhrenreichCohen59},
which are then converted into the conductivities $\bar\sigma_{zz}$
by means of the Drude relation. Electronic dissipation processes
are included in our calculations within the relaxation-time
approximation (electron scattering length of $130R_{CN}$ was
used~\cite{Lazzeri05}). We did not include excitonic many-electron
correlations, however, as they mostly affect the real conductivity
$\mbox{Re}(\bar\sigma_{zz})$ which is responsible for the CN
optical absorption~\cite{Pedersen04,Spataru04,Ando}, whereas we
are interested here in $\mbox{Re}(1/\bar\sigma_{zz})$ representing
the surface plasmon DOS according to Eq.~(\ref{plDOS}). This
function is only non-zero when the two conditions,
$\mbox{Im}[\bar\sigma_{zz}(x)]=0$ and
$\mbox{Re}[\bar\sigma_{zz}(x)]\rightarrow0$, are fulfilled
simultaneously~\cite{Kempa02,Kempa89,LinShung97}. These result in
the peak structure of the function $\mbox{Re}(1/\bar\sigma_{zz})$
as is seen in Fig.~\ref{fig3}. It~is~also seen from the comparison
of Fig.~\ref{fig3}~(b) with Fig.~\ref{fig3}~(a) that the peaks
broaden as the CN diameter decreases. This is consistent with the
stronger hybridization effects in smaller-diameter
CNs~\cite{Blase94}.

Left panels in Figs.~\ref{fig4}(a) and \ref{fig4}(b) show the
lowest-energy plasmon DOS resonances calculated for the (11,0) and
(10,0) CNs as given by the function $\rho(x)$ in
Eq.~(\ref{plDOS}). Also shown there are the corresponding
fragments of the functions $\mbox{Re}[\bar\sigma_{zz}(x)]$ and
$\mbox{Im}[\bar\sigma_{zz}(x)]$. In all graphs the lower
dimensionless energy limits are set up to be equal to the lowest
bright exciton excitation energy [$E^{(11)}_{exc}=1.21$~eV
($x=0.224$) and $1.00$~eV ($x=0.185$) for the (11,0) and (10,0)
CN, respectively, as reported in Ref.\cite{Spataru05} by directly
solving the Bethe-Salpeter equation]. Peaks in $\rho(x)$ are seen
to coincide in energy with zeros of
$\mbox{Im}[\bar\sigma_{zz}(x)]$ \{or zeros of
$\mbox{Re}[\epsilon_{zz}(x)]$\}, clearly indicating the plasmonic
nature of the CN surface excitations under
consideration~\cite{Kempa02,Kempa}. They describe the surface
plasmon modes associated with the transversely quantized interband
electronic transitions in CNs~\cite{Kempa02}. As is seen in
Fig.~\ref{fig4} (and in Fig.~\ref{fig3}), the interband plasmon
excitations occur in CNs slightly above the first bright exciton
excitation energy~\cite{Ando}, in the frequency domain where the
imaginary conductivity (or the real dielectric function) changes
its sign. This is a unique feature of the complex dielectric
response function, the consequence of the general
Kramers-Kr\"{o}nig relation~\cite{VogelWelsch}.

We further take advantage of the sharp peak structure of $\rho(x)$
and solve the dispersion equation~(\ref{dispeq}) for $x_\mu$
analytically using the Lorentzian approximation
\begin{equation}
\rho(x)\!\approx\!\frac{\rho(x_p)\Delta x_{p}^2}
{(x-x_{p})^2+\Delta x_{p}^2}\,. \label{rhox}
\end{equation}
Here, $x_{p}$ and $\Delta x_{p}$ are, respectively, the position
and the half-width-at-half-maximum of the plasmon resonance
closest to the lowest bright exciton excitation energy in the same
nanotube (as shown in the left panels of Fig.~\ref{fig4}). The
integral in Eq.~(\ref{dispeq}) then simplifies to the form
\[
\frac{2}{\pi}\int_0^\infty\!\!\!\!\!dx\,\frac{x\,\bar\Gamma_0^f(x)\rho(x)}{x_\mu^2-x^2}
\approx\frac{F(x_p)\Delta x_{p}^2}{x_\mu^2-x_p^2}\!
\int_0^\infty\!\!\!\!\!\frac{dx}{(x-x_{p})^2+\Delta x_{p}^2}
\]\vspace{-0.25cm}
\[
=\frac{F(x_p)\Delta x_{p}}{x_\mu^2-x_p^2}
\left[\arctan\!\left(\frac{x_p}{\Delta x_p}\right)+
\frac{\pi}{2}\right]
\]
with $F(x_p)=2x_p\bar\Gamma_0^f(x_p)\rho(x_p)/\pi$. This
expression is valid for all $x_\mu$ apart from those located in
the narrow interval $(x_p-\Delta x_p,x_p+\Delta x_p)$ in the
vicinity of the plasmon resonance, provided that the resonance is
sharp enough. Then, the dispersion equation becomes the
biquadratic equation for $x_\mu$ with the following two positive
solutions (the dispersion curves) of interest to us

\begin{equation}
x_{1,2}=\sqrt{\frac{\varepsilon_f^2+x_{p}^2}{2}\pm\frac{1}{2}
\sqrt{(\varepsilon_f^2\!-x_{p}^2)^2+F_{\!p}\,\varepsilon_f}}\,.
\label{dispsol}
\end{equation}
Here, $F_{\!p}=4F(x_p)\Delta x_p(\pi-\Delta x_{p}/x_{p})$ with the
$\arctan$-function expanded to linear terms in $\Delta
x_p/x_p\ll1$.

The dispersion curves (\ref{dispsol}) are shown in the right
panels in Figs.~\ref{fig4}(a) and \ref{fig4}(b) as functions of
the dimensionless longitudinal quasi-momentum.~In these
calculations, we estimated the interband transition matrix element
in $\bar\Gamma_0^f(x_p)$ [Eq.~(\ref{Gamma0f})] from the equation
$|d_f|^2=3\hbar\lambda^3/4\tau_{ex}^{rad}$ according to Hanamura's
general theory of the exciton radiative decay in spatially
confined systems~\cite{Hanamura}, where $\tau_{ex}^{rad}$ is the
exciton intrinsic radiative lifetime, and $\lambda=2\pi c\hbar/E$
with $E$ being the exciton total energy given in our case by
Eq.~(\ref{Ef}).~For zigzag-type CNs considered here, the first
Brillouin zone of the longitudinal quasi-momentum is given by
$-2\pi\hbar/3b\le\hbar k_z\le2\pi\hbar/3b$~\cite{Dresselhaus,Dai}.
The total energy of the ground-internal-state exciton can then be
written as $E=E_{exc}+(2\pi\hbar/3b)^2t^2/2M_{ex}$ with $-1\le
t\le1$ representing the dimensionless longitudinal quasi-momentum.
In our calculations we used the lowest bright exciton parameters
$E^{(11)}_{exc}=1.21$~eV and $1.00$~eV, $\tau_{ex}^{rad}=14.3$~ps
and $19.1$~ps, $M_{ex}=0.44m_0$ and $0.19m_0$ ($m_0$ is the
free-electron mass) for the (11,0) CN and (10,0) CN, respectively,
as reported in Ref.\cite{Spataru05} by directly solving the
Bethe-Salpeter equation.

Both graphs in the right panels in Fig.~\ref{fig4} are seen to
demonstrate a clear anticrossing behavior with the (Rabi) energy
splitting $\sim\!0.1$~eV. This indicates the formation of the
strongly coupled surface plasmon-exciton excitations in the
nanotubes under consideration. It is important to realize that
here we deal with the strong exciton-plasmon interaction supported
by an individual quasi-1D nanostructure --- a single-walled
(small-diameter) semiconducting carbon nanotube, as opposed to the
artificially fabricated metal-semiconductor nanostructures studied
previuosly~\cite{Bellessa,Govorov,Fedutik} where the metallic
component normally carries the plasmon and the semiconducting one
carries the exciton. It is also important that the effect comes
not only from the height but also from the width of the plasmon
resonance as it is seen from the definition of the $F_p$ factor in
Eq.~(\ref{dispsol}). In other words, as long as the plasmon
resonance is sharp enough (which is always the case for interband
plasmons), so that the Lorentzian approximation (\ref{rhox})
applies, the effect is determined by the area under the plasmon
peak in the DOS function~(\ref{plDOS}) rather than by the peak
height as one would expect.

However, the formation of the strongly coupled exci\-ton-plasmon
states is only possible if the exciton total energy is in
resonance with the energy of a surface plasmon mode. The exciton
energy can be tuned to the nearest plasmon resonance in ways used
for excitons in semiconductor quantum microcavities
--- thermally~\cite{Reithmaier,Yoshie,Peter} (by elevating sample temperature),
or/and electrostatically~\cite{MillerPRL,Miller,Zrenner,Krenner}
(via the quantum confined Stark effect with an external
electrostatic field applied perpendicular to the CN axis). As is
seen from Eqs.~(\ref{Ef}) and (\ref{Eexcf}), the two possibilities
influence the different degrees of freedom of the quasi-1D exciton
--- the (longitudinal) kinetic energy and the excitation energy,
respectively. Below we study the (less trivial) electrostatic
field effect on the exciton excitation energy in carbon nanotubes.

\subsection{The perpendicular electrostatic field effect}

The optical properties of semiconducting CNs in an external
electrostatic field directed along the nanotube axis were studied
theoretically in Ref.~\cite{Perebeinos07}. Strong oscillations in
the band-to-band absorption and the quadratic Stark shift of the
exciton absorption peaks with the field increase, as well as the
strong field dependence of the exciton ionization rate, were
predicted for CNs of different diameters and chiralities. Here, we
focus on the perpendicular electrostatic field orientation.~We
study how the electrostatic field applied perpendicular to the CN
axis affects the CN band gap, the exciton binding/excitation
energy, and the interband surface plasmon energy, to explore the
tunability of the strong exciton-plasmon coupling effect predicted
above.~The problem is similar to the well-known quantum confined
Stark effect first observed for the excitons in semiconductor
quantum wells~\cite{MillerPRL,Miller}. However, the cylindrical
surface symmetry of the excitonic states brings new peculiarities
to the quantum confined Stark effect in CNs. In what follows we
will generally be interested only in the lowest internal energy
(ground) excitonic state, and so the internal state index $f$ in
Eqs.~(\ref{Ef}) and (\ref{Eexcf}) will be omitted for brevity.

Because the nanotube is modelled by a continuous, infinitely thin,
anisotropically conducting cylinder in our macroscopic QED
approach, the actual local symmetry of the excitonic wave function
resulted from the graphene Brillouin zone structure is disregarded
in our model (see, e.g., reviews~\cite{Dresselhaus07,Ando}). The
local symmetry is implicitly present in the surface axial
conductivity though, which we calculate beforehand as described
above~\cite{EndNote1}.

We start with the Schr\"{o}dinger equation for the electron
located at $\textbf{r}_{e}=\{R_{CN},\varphi_{e},z_{e}\}$ and the
hole located at $\textbf{r}_{h}=\{R_{CN},\varphi_{h},z_{h}\}$ on
the nanotube surface. They interact with each other through the
Coulomb potential
$V(\textbf{r}_e,\textbf{r}_h)=-e^2/\epsilon|\textbf{r}_e\!-\textbf{r}_h|$,
where $\epsilon=\epsilon_{zz}(0)$.~The external electrostatic
field $\textbf{F}=\{F,0,0\}$ is directed perpendicular to the CN
axis (along the $x$-axis in Fig.~\ref{fig1}). The Schr\"{o}dinger
equation is of the form
\begin{equation}
\left[\hat{H}_{e}(\textbf{F})+\hat{H}_{h}(\textbf{F})+
V(\textbf{r}_e,\textbf{r}_h)\right]\!
\Psi(\textbf{r}_e,\textbf{r}_h)=E\Psi(\textbf{r}_e,\textbf{r}_h)
\label{Schroedinger}
\end{equation}
with
\begin{equation}
\hat{H}_{e,h}(\textbf{F})=-\frac{\hbar^2}{2m_{e,h}}\left(\!
\frac{1}{R_{CN}^2}\frac{\partial^2}{\partial\varphi^2_{e,h}}+
\frac{\partial^2}{\partial z^2_{e,h}}\right)\!\mp
e\textbf{r}_{e,h}\!\cdot\textbf{F}\label{Heh}
\end{equation}

We further separate out the translational and relative degrees of
freedom of the electron-hole pair by transforming the longitudinal
(along the CN axis) motion of the pair into its center-of-mass
coordinates given by $Z=(m_ez_e+m_hz_h)/M_{ex}$ and $z=z_e-z_h$.
The exciton wave function is approximated as follows
\begin{equation}
\Psi(\textbf{r}_e,\textbf{r}_h)=e^{ik_zZ}\phi_{ex}(z)\psi_e(\varphi_e)\psi_h(\varphi_h).
\label{wfunceh}
\end{equation}
The complex exponential describes the exciton center-of-mass
motion with the longitudinal quasi-momentum $k_z$ along the CN
axis. The function $\phi_{ex}(z)$ represents the longitudinal
relative motion of the electron and the hole inside the exciton.
The functions $\psi_e(\varphi_e)$ and $\psi_h(\varphi_h)$ are the
electron and hole subband wave functions, respectively, which
represent their confined motion along the circumference of the
cylindrical nanotube surface.

Each of the functions is assumed to be normalized to unity.
Equations~(\ref{Schroedinger}) and (\ref{Heh}) are then rewritten
in view of Eqs.~(\ref{Ef})--(\ref{Egkfi}) to yield
\begin{equation}
\left[-\frac{\hbar^2}{2m_eR_{CN}^2}\frac{\partial^2}{\partial\varphi^2_e}-
eR_{CN}F\cos(\varphi_e)\right]\!\psi_e(\varphi_e)=\varepsilon_{e}\psi_e(\varphi_e),
\label{wfe}
\end{equation}
\begin{equation}
\left[-\frac{\hbar^2}{2m_hR_{CN}^2}\frac{\partial^2}{\partial\varphi^2_h}+
eR_{CN}F\cos(\varphi_h)\right]\!\psi_h(\varphi_h)\!=\!\varepsilon_h\psi_h(\varphi_h)\!,
\label{wfh}
\end{equation}
\begin{equation}
\left[-\frac{\hbar^2}{2\mu}\frac{\partial^2}{\partial\,\!z^2}+
V_{\mbox{\small{eff}}}(z)\right]\!\phi_{ex}(z)=E_b\phi_{ex}(z),
\label{wfexc}
\end{equation}
where $\mu=m_em_h/M_{ex}$ is the exciton reduced mass, and
$V_{\mbox{\small eff}}$ is the effective longitudinal
electron-hole Coulomb interaction potential given by
\begin{equation}
V_{\mbox{\small
eff}}(z)=\!-\frac{e^2}{\epsilon}\!\int_0^{2\pi}\!\!\!\!\!\!d\varphi_e\!\!\int_0^{2\pi}\!\!\!\!\!\!
d\varphi_h|\psi_e(\varphi_e)|^2|\psi_h(\varphi_h)|^2V(\varphi_e,\varphi_h,z)
\label{Veff}
\end{equation}
with $V$ being the original electron-hole Coulomb potential
written in the cylindrical coordinates as
\begin{equation} V(\varphi_e,\varphi_h,z)=\!
\frac{1}{\{z^2+4R_{CN}^2\sin^2[(\varphi_{e}\!-\varphi_{h})/2]\}^{1/2}}.
\label{V}
\end{equation}
The exciton problem is now reduced to the 1D equation
(\ref{wfexc}), where the exciton binding energy does depend on the
perpendicular electrostatic field through the electron and hole
subband functions $\psi_{e,h}$ given by the solutions of
Eqs.~(\ref{wfe}) and (\ref{wfh}) and entering the effective
electron-hole Coulomb interaction potential (\ref{Veff}).

The set of Eqs.~(\ref{wfe})-(\ref{V}) is analyzed in
Appendix~\ref{appD}. One of the main results obtained in there is
that the effective Coulomb potential (\ref{Veff}) can be
approximated by an attractive cusp-type cutoff potential of the
form
\begin{equation}
V_{\mbox{\small
eff}}(z)\approx-\frac{e^2}{\epsilon[|z|+z_0(j,F)]},
\label{Vcutoff}
\end{equation}
where the cutoff parameter $z_0$ depends on the perpendicular
electrostatic field strength and on the electron-hole azimuthal
transverse quantization index $j=1,2,...$ (excitons are created in
interband transitions involving valence and conduction subbands
with the same quantization index~\cite{Ando} as shown in
Fig.~\ref{fig2}). Specifically,
\begin{equation}
z_0(j,F)\approx2R_{CN}\frac{\pi-2\ln2\,[1-\Delta_j(F)]}{\pi+2\ln2\,[1-\Delta_j(F)]}
\label{z0}
\end{equation}
with $\Delta_j(F)$ given to the second order approximation in the
electric field by
\begin{eqnarray}
\Delta_j(F)&\!\!\approx\!\!&2\mu M_{ex}\frac{e^2R_{CN}^6w^2_j}{\hbar^4}\,F^2,\label{DeltaF}\\
w_j&\!\!=\!\!&\frac{\theta(j\!-\!2)}{1-2j}+\frac{1}{1+2j},\nonumber
\end{eqnarray}
where $\theta(x)$ is the unit step function. Approximation
(\ref{Vcutoff}) is formally valid when $z_0(j,F)$ is much less
than the exciton Bohr radius $a_B^\ast$ $(=\epsilon\hbar^2/\mu
e^2)$ which is estimated to be $\sim\!10R_{CN}$ for the first
($j\!=\!1$ in our notations here) exciton in
CNs~\cite{Pedersen03}. As is seen from Eqs.~(\ref{z0}) and
(\ref{DeltaF}), this is always the case for the first exciton for
those fields where the perturbation theory applies, i.~e. when
$\Delta_1(F)<1$ in Eq.~(\ref{DeltaF}).

Equation~(\ref{wfexc}) with the potential (\ref{Vcutoff}) formally
coincides with the one studied by Ogawa and Takagahara in their
treatments of excitonic effects in 1D semiconductors with no
external electrostatic field applied~\cite{Takagahara}.~The only
difference in our case is that our cutoff parameter (\ref{z0}) is
field dependent. We therefore follow Ref.~\cite{Takagahara} and
find the ground-state binding energy $E^{(11)}_b$ for the first
exciton we are interested in here from the transcendental equation
\begin{equation}
\ln\!\left[\frac{2z_0(1,F)}{\hbar}\sqrt{2\mu|E^{(11)}_b|}\,\right]
+\frac{1}{2}\sqrt{\frac{|E^{(11)}_b|}{Ry^\ast}}=0. \label{Ebnum}
\end{equation}
In doing so, we first find the exciton Rydberg energy,~$Ry^\ast$
$(=\!\mu e^4/2\hbar^2\epsilon^2)$, from this equation at
$F\!=\!0$. We use the diameter- and chirality-dependent electron
and hole effective masses from Ref.~\cite{Jorio}, and the first
bright exciton binding energy of 0.76~eV for both (11,0) and
(10,0) CN as reported in Ref.~\cite{Capaz} from \emph{ab initio}
calculations. We obtain $Ry^\ast=4.02$~eV and $0.57$~eV for
the~(11,0) tube and (10,0) tube, respectively. The difference of
about one order of magnitude reflects the fact that these are the
semiconducting CNs of different types --- type-I and type-II,
respectively, based on $(2n+m)$ families~\cite{Jorio}. The
parameters $Ry^\ast$ thus obtained are then used to find
$|E^{(11)}_b|$ as functions of $F$ by numerically solving
Eq.~(\ref{Ebnum}) with $z_0(1,F)$ given by Eqs.~(\ref{z0}) and
(\ref{DeltaF}).

\begin{figure}[t]
\epsfxsize=8.65cm\centering{\epsfbox{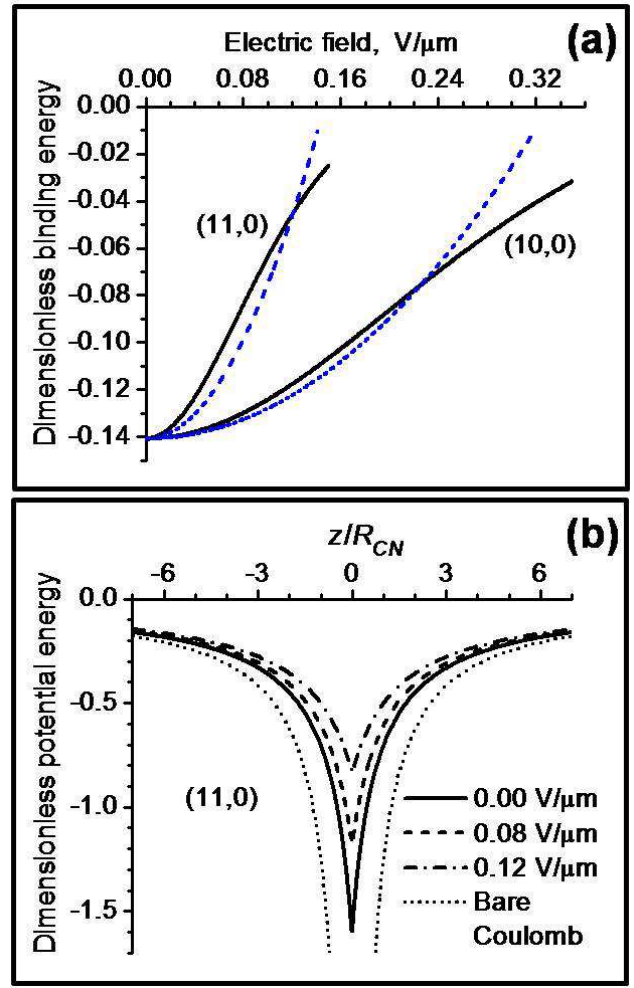}}\caption{(Color
online)~(a)~Calculated binding energies of the first bright
exciton in the (11,0) and (10,0) CNs as functions of the
perpendicular electrostatic field applied.~Solid lines are the
numerical solutions to Eq.~(\ref{Ebnum}), dashed lines are the
quadratic approximations as given by Eq.~(\ref{Ebapprox}).
(b)~Field dependence of the effective cutoff Coulomb potential
(\ref{Vcutoff}) in the (11,0) CN. The dimensionless energy is
defined as [\emph{Energy}]/$2\gamma_0$, according to
Eq.~(\ref{dimless}).} \label{fig5}
\end{figure}

The calculated (negative) binding energies are shown by the solid
lines in Fig.~\ref{fig5}(a). Also shown there by dashed lines are
the functions
\begin{equation}
E^{(11)}_b(F)\approx E^{(11)}_b[1-\Delta_1(F)] \label{Ebapprox}
\end{equation}
with $\Delta_1(F)$ given by Eq.~(\ref{DeltaF}). They are seen to
be fairly good analytical (quadratic in field) approximations to
the numerical solutions of Eq.~(\ref{Ebnum}) in the range of not
too large fields. The exciton binding energy decreases very
rapidly in its absolute value as the field increases. Fields of
only $\sim\!0.1-0.2$~V/$\mu$m are required to decrease
$|E^{(11)}_b|$ by a factor of $\sim\!2$ for the CNs considered
here. The reason is the perpendicular field shifts up the "bottom"
of the effective potential (\ref{Vcutoff}) as shown in
Fig.~\ref{fig5}(b) for the (11,0) CN. This makes the potential
shallower and pushes bound excitonic levels up, thereby decreasing
the exciton binding energy in its absolute value. As this takes
place, the shape of the potential does not change, and the
longitudinal relative electron-hole motion remains finite at all
times.~As a consequence, no tunnel exciton ionization occurs in
the perpendicular field, as opposed to the longitudinal
electrostatic field (Franz-Keldysh) effect studied in
Ref.~\cite{Perebeinos07} where the non-zero field creates the
potential barrier separating out the regions of finite and
infinite relative motion and the exciton becomes ionized as the
electron tunnels to infinity.

The binding energy is only the part of the exciton excitation
energy (\ref{Eexcf}). Another part comes from the band gap energy
(\ref{Egkfi}), where $\varepsilon_{e}$ and $\varepsilon_{h}$ are
given by the solutions of Eqs.~(\ref{wfe}) and (\ref{wfh}),
respectively. Solving them to the leading (second) order
perturbation theory approximation in the field
(Appendix~\ref{appD}), one obtains
\begin{equation}
E^{(jj)}_g(F)\approx
E^{(jj)}_g\!\left[1-\frac{m_e\Delta_j(F)}{2M_{ex}j^2w_j}-\frac{m_h\Delta_j(F)}{2M_{ex}j^2w_j}\right]\!,
\label{EgF}
\end{equation}
where the electron and hole subband shifts are written separately.
This, in view of Eq.~(\ref{DeltaF}), yields the first band gap
field dependence in the form
\begin{equation}
E^{(11)}_g(F)\approx
E^{(11)}_g\!\left[1-\frac{3}{2}\,\Delta_1(F)\right]\!,
\label{EgF11}
\end{equation}
The bang gap decrease with the field in Eq.~(\ref{EgF11}) is
stronger than the opposite effect in the negative exciton binding
energy given (to the same order approximation in field) by
Eq.~(\ref{Ebapprox}). Thus, the first exciton excitation energy
(\ref{Eexcf}) will be gradually decreasing as the perpendicular
field increases, shifting the exciton absorption peak to the red.
This is the basic feature of the quantum confined Stark effect
observed previously in semiconductor
nanomaterials~\cite{MillerPRL,Miller,Zrenner,Krenner}. The field
dependences of the higher interband transitions exciton excitation
energies are suppressed by the rapidly (quadratically) increasing
azimuthal quantization numbers in the denominators of
Eqs.~(\ref{DeltaF}) and (\ref{EgF}).

Lastly, the perpendicular field dependence of the interband
plasmon resonances can be obtained from the frequency dependence
of the axial surface conductivity due to excitons (see
Ref.~\cite{Ando} and refs. therein). One has
\begin{equation}
\sigma^{ex}_{zz}(\omega)\sim\!\!\!\sum_{j=1,2,...}\!\frac{-i\hbar\omega
f_j}{[E^{(jj)}_{exc}]^2\!-(\hbar\omega)^2\!-2i\hbar^2\omega/\tau},
\label{Ando}
\end{equation}
where $f_j$ and $\tau$ are the exciton oscillator strength and
relaxation time, respectively. The plasmon frequencies are those
at which the function $\mbox{Re}[1/\sigma^{ex}_{zz}(\omega)]$ has
maxima. Testing it for maximum in the domain
$E^{(11)}_{exc}\!\!<\!\hbar\omega<\!E^{(22)}_{exc}$, one finds the
first interband plasmon resonance energy to be (in the limit
$\tau\!\rightarrow\!\infty$)
\begin{equation}
E^{(11)}_p=\sqrt{\frac{[E^{(11)}_{exc}]^2+[E^{(22)}_{exc}]^2}{2}}\,.
\label{Epl}
\end{equation}
Using the field dependent $E^{(11)}_{exc}$ given by
Eqs.~(\ref{Eexcf}),~(\ref{Ebapprox}) and (\ref{EgF11}), and
neglecting the field dependence of $E^{(22)}_{exc}$, one obtains
to the second order approximation in the field
\begin{equation}
E^{(11)}_p(F)\approx
E^{(11)}_p\!\left[1-\frac{1+\!E^{(11)}_{g}\!/2E^{(11)}_{exc}}
{1+\!E^{(22)}_{exc}\!/E^{(11)}_{exc}}\;\Delta_1(F)\right]\!.
\label{EplF}
\end{equation}

\begin{figure}[t]
\epsfxsize=8.65cm\centering{\epsfbox{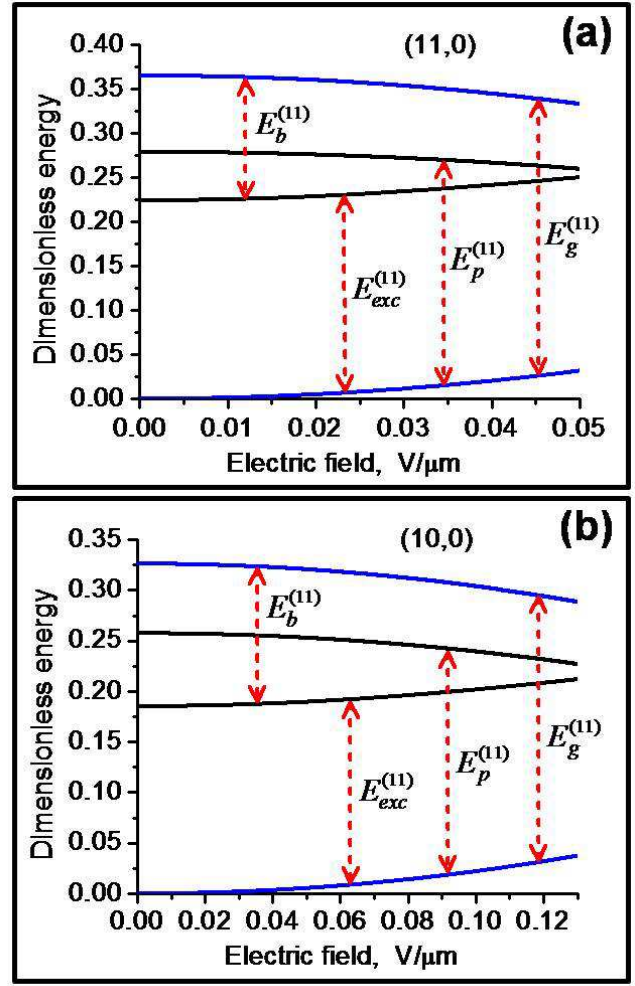}}\caption{(Color
online)~(a),(b)~Calculated dependences of the first bright exciton
parameters in the (11,0) and (10,0) CNs, respectively, on the
electrostatic field applied perpendicular to the nanotube axis.
The dimensionless energy is defined as
[\emph{Energy}]/$2\gamma_0$, according to Eq.~(\ref{dimless}). The
energy is measured from the top of the first unperturbed hole
subband.} \label{fig6}
\end{figure}

Figure~\ref{fig6} shows the results of our calculations of the
field dependences for the first bright exciton parameters in the
(11,0) and (10,0) CNs. The energy is measured from the top of the
first unperturbed hole subband (as shown in Fig.~\ref{fig2}, right
panel). The binding energy field dependence was calculated
numerically from Eq.~(\ref{Ebnum}) as described above [shown in
Fig.~\ref{fig5}~(a)]. The band gap field dependence and the
plasmon energy field dependence were calculated from
Eqs.~(\ref{EgF}) and (\ref{EplF}), respectively. The zero-field
excitation energies and zero-field binding energies were taken to
be those reported in Ref.~\cite{Spataru05} and in
Ref.~\cite{Capaz}, respectively, and we used the diameter- and
chirality-dependent electron and hole effective masses from
Ref.~\cite{Jorio}. As is seen in Fig.~\ref{fig6}~(a)~and (b), the
exciton excitation energy and the interband plasmon energy
experience red shift in both nanotubes as the field increases.
However, the excitation energy red shift is very small (barely
seen in the figures) due to the negative field dependent
contribution from the exciton binding energy.~So,
$E^{(11)}_{exc}(F)$ and $E^{(11)}_{p}(F)$ approach each other as
the field increases, thereby bringing the total exciton energy
(\ref{Ef}) in resonance with the surface plasmon mode due to the
non-zero longitudinal kinetic energy term at finite
temperature~\cite{EndNote2}. Thus, the electrostatic field applied
perpendicular to the CN axis (the quantum confined Stark effect)
may be used to tune the exciton energy to the nearest interband
plasmon resonance, to put the exciton-surface plasmon interaction
in small-diameter semiconducting CNs to the strong-coupling
regime.

\subsection{The optical absorption}

Here we analyze the longitudinal exciton absorption line shape as
its energy is tuned to the nearest interband surface plasmon
resonance. Only longitudinal excitons (excited by light polarized
along the CN axis) couple to the surface plasmon modes as
discussed at the very beginning of this section (see
Ref.~\cite{UryuAndo} for the perpendicular light exciton
absorption in CNs). We follow the optical absorption/emission
lineshape theory developed recently for atomically doped
CNs~\cite{Bondarev06}. (Obviously, the absorption line shape
coincides with the emission line shape if the monochromatic
incident light beam is used in the absorption experiment.) When
the $f$-internal state exciton is excited and the nanotube's
surface EM field subsystem is in vacuum state, the time-dependent
wave function of the whole system "exciton+field" is of the
form~\cite{EndNote1}
\begin{eqnarray}
|\psi(t)\rangle&=&\sum_{{\bf
k},f}C_{f}(\textbf{k},t)\,e^{-i\tilde{E}_{f}({\bf k})t/\hbar}
|\{1_f(\textbf{k})\}\rangle_{ex}|\{0\}\rangle\label{wfunc}\\
&+&\sum_{\bf k}\int_{0}^{\infty}\!\!\!\!\!\!d\omega\,
C(\textbf{k},\omega,t)\,e^{-i\omega t}
|\{0\}\rangle_{ex}|\{1(\mathbf{k},\omega)\}\rangle.\nonumber
\end{eqnarray}
Here, $|\{1_f(\textbf{k})\}\rangle_{ex}$ is the excited
single-quantum Fock state with one exciton and
$|\{1(\textbf{k},\omega)\}\rangle$ is that with one surface
photon. The vacuum states are $|\{0\}\rangle_{ex}$ and
$|\{0\}\rangle$ for the exciton subsystem and field subsystem,
respectively. The coefficients $C_{f}(\textbf{k},t)$ and
$C(\textbf{k},\omega,t)$ stand for the population probability
amplitudes of the respective states of the whole system. The
exciton energy is of the form
$\tilde{E}_{f}(\textbf{k})\!=\!E_f(\textbf{k})\!-i\hbar/\tau$ with
$E_f(\textbf{k})$ given by Eq.~(\ref{Ef}) and $\tau$ being the
phenomenological exciton relaxation time constant [assumed to be
such that $\hbar/\tau\!\ll\!E_{f}(\textbf{k})$] to account for
other possible exciton relaxation processes. From the literature
we have $\tau_{ph}\sim30\!-\!100$~fs for the exciton-phonon
scattering~\cite{Perebeinos07}, $\tau_d\sim50$~ps for the exciton
scattering by defects~\cite{Hagen05,Prezhdo08}, and
$\tau_{rad}\sim10\,\mbox{ps}-10\,\mbox{ns}$ for the radiative
decay of excitons~\cite{Spataru05}. Thus, the scattering by
phonons is the most likely exciton relaxation mechanism.

We transform the total Hamiltonian~(\ref{Htot})--(\ref{Hint}) to
the $\mathbf{k}$-representation using Eqs.~(\ref{Bkf}) and
(\ref{orthog}) (see Appendix~\ref{appA}), and apply it to the wave
function in Eq.~(\ref{wfunc}). We obtain the following set of the
two simultaneous differential equations for the coefficients
$C_{f}(\textbf{k},t)$ and $C(\textbf{k},\omega,t)$ from the time
dependent Schr\"{o}dinger equation
\begin{eqnarray}
\mbox{\it\.C}_{f}(\textbf{k},t)\,e^{-i\tilde{E}_{f}({\bf\!\,k})t/\hbar}\!\!&=&\!\!
-\frac{i}{\hbar}\sum_{{\bf\!\,k^\prime}}\int_{0}^{\infty}\!\!\!\!\!d\omega\,
\mbox{g}^{(+)}_f(\mathbf{k},\mathbf{k}^\prime\!,\omega)\hspace{1cm}
\label{paexc}\\
&\times&\!C(\textbf{k}^\prime\!,\omega,t)\,e^{-i\omega\!\,t},\nonumber
\end{eqnarray}
\vspace{-0.5cm}
\begin{eqnarray}
\mbox{\it\.C\,}(\textbf{k}^\prime\!,\omega,t)\,e^{-i\omega\!\,t}\delta_{{\bf\!\,k}{\bf\!\,k}^\prime}\!\!&=&\!\!
-\frac{i}{\hbar}\sum_f\,[\mbox{g}^{(+)}_f(\mathbf{k},\mathbf{k}^\prime\!,\omega)]^\ast\hspace{1cm}
\label{papho}\\
&\times&\!C_f(\textbf{k},t)\,e^{-i\tilde{E}_{f}({\bf\!\,k})t/\hbar}.\nonumber
\end{eqnarray}
The $\delta$-symbol on the left in Eq.~(\ref{papho}) ensures that
the momentum conservation is fulfilled in the exciton-photon
transitions, so that the annihilating exciton creates the surface
photon with the same momentum and vice versa. In terms of the
probability amplitudes above, the exciton emission intensity
distribution is given by the final state probability at very long
times corresponding to the complete decay of all initially excited
excitons,
\begin{eqnarray}
I(\omega)\!&=&\!|C(\mathbf{k},\omega,t\!\rightarrow\!\infty)|^{2}=
\frac{1}{\hbar^2}\sum_f|\mbox{g}^{(+)}_f(\mathbf{k},\mathbf{k},\omega)|^2\nonumber\\
&\times&\!\left|\int_0^\infty\!\!\!\!\!dt^\prime\,C_f(\textbf{k},t^\prime)\,
e^{-i[\tilde{E}_{f}({\bf\!\,k})-\hbar\omega]t^\prime\!/\hbar}\right|^2\!.\label{Iomega}
\end{eqnarray}
Here, the second equation is obtained by the formal integration of
Eq.~(\ref{papho}) over time under the initial condition
$C\,(\textbf{k},\omega,0)\!=\!0$. The emission intensity
distribution is thus related to the exciton population probability
amplitude $C_f(\mathbf{k},t)$ to be found from Eq.~(\ref{paexc}).

The set of simultaneous equations (\ref{paexc}) and (\ref{papho})
[and Eq.~(\ref{Iomega}), respectively] contains no approximations
except the (commonly used) neglect of many-particle excitations in
the wave function (\ref{wfunc}).~We now apply these equations to
the exciton-surface-plasmon system in small-diameter
semiconducting CNs. The interaction matrix element~in
Eqs.~(\ref{paexc}) and (\ref{papho}) is then given by the
$\mathbf{k}$-transform of Eq.~(\ref{gfpar}), and has the following
property (Appendix~\ref{appC})
\begin{equation}
\frac{1}{2\gamma_0\hbar}\,|\mbox{g}^{(+)}_f(\mathbf{k},\mathbf{k},\omega)|^2
=\frac{1}{2\pi}\,\bar\Gamma_0^f(x)\rho(x) \label{gpkk}
\end{equation}
with $\bar\Gamma_0^f(x)$ and $\rho(x)$ given by
Eqs.~(\ref{Gamma0f}) and (\ref{plDOS}), respectively. We further
substitute the result of the formal integration of
Eq.~(\ref{papho}) [with $C\,(\textbf{k},\omega,0)\!=\!0$] into
Eq.~(\ref{paexc}), use Eq.~(\ref{gpkk}) with $\rho(x)$
approximated by the Lorentzian~(\ref{rhox}), calculate the
integral over frequency analytically, and differentiate the result
over time to obtain the following second order ordinary
differential equation for the exciton probability amplitude
[dimensionless variables, Eq.~(\ref{dimless})]
\[
\mbox{\it\"{C}}_f(\beta)+[\Delta
x_p-\Delta\varepsilon_f+i(x_p-\varepsilon_f)]\mbox{\it\.{C}}_f(\beta)+(X_f/2)^{2}C_f(\beta)\!=\!0,
\]
where $X_f\!=\![2\Delta x_{p}\bar\Gamma_f(x_p)]^{1/2}$ with
$\bar\Gamma_f(x_{p})\!=\!\bar\Gamma_0^f(x_{p})\rho(x_{p})$,
$\Delta\varepsilon_f=\hbar/2\gamma_0\tau$,
$\beta=2\gamma_0t/\hbar$ is the dimensionless time,~and the
\textbf{k}-dependence is omitted for brevity. When the total
exciton energy is close to a plasmon resonance,
$\varepsilon_f\!\approx\!x_{p}$, the solution of this equation is
easily found to be
\begin{eqnarray}
C_f(\beta)\!\!\!&\approx&\!\!\!\frac{1}{2}\left(\!1+\frac{\delta
x}{\sqrt{\delta x^2-X_f^2}}\right)e^{-\left(\delta
x-\sqrt{\delta x^2-X_f^2}\right)\beta/2}\hskip0.5cm\label{Cfapp}\\
&+&\!\!\!\frac{1}{2}\left(\!1-\frac{\delta x}{\sqrt{\delta
x^2-X_f^2}}\right)e^{-\left(\delta x+\sqrt{\delta
x^2-X_f^2}\right)\beta/2},\nonumber
\end{eqnarray}
where $\delta x=\Delta x_p-\Delta\varepsilon_f>0$ and
$X_f\!=\![2\Delta x_{p}\bar\Gamma_f(\varepsilon_f)]^{1/2}$. This
solution is valid when $\varepsilon_f\!\approx\!x_p$ regardless
of~the strength of the exciton-surface-plasmon coupling.~It yields
the exponential decay of the excitons into plasmons,
$|C_f(\beta)|^2\!\approx\!\exp[-\bar\Gamma_f(\varepsilon_f)\beta]$,
in the weak coupling regime where the coupling parameter
$(X_f/\delta x)^{2}\!\ll\!1$. If, on the other hand, $(X_f/\delta
x)^{2}\!\gg\!1$,~then the strong coupling regime occurs, and the
decay of the excitons into plasmons proceeds via damped Rabi
oscillations, $|C_f(\beta)|^{2}\!\approx\!\exp(-\delta
x\beta)\cos^{2}(X_f\beta/2)$.~This is very similar to what was
earlier reported for an excited two-level atom near the nanotube
surface~\cite{Bondarev02,Bondarev04,Bondarev04pla,Bondarev06trends}.~Note,
however, that here we have the exciton-phonon scattering as well,
which facilitates the strong exciton-plasmon coupling by
decreasing $\delta x$ in the coupling parameter.~In other words,
the phonon scattering broadens the (longitudinal) exciton momentum
distribution~\cite{Bondarev05Ps}, thus effectively increasing the
fraction of the excitons with $\varepsilon_f\!\approx\!x_p$.

In view of Eqs.~(\ref{gpkk}) and (\ref{Cfapp}), the exciton
emission intensity (\ref{Iomega}) in the vicinity of the plasmon
resonance takes the following (dimensionless) form

\begin{equation}
\bar{I}(x)\approx\bar{I}_{0}(\varepsilon_f)\sum_f\left|\int_{0}^{\infty}\!\!\!\!\!\!d\beta\,C_{f}(\beta)\,
e^{i(x-\varepsilon_f+i\Delta\varepsilon_f)\beta}\right|^{2},
\label{Ix}
\end{equation}
where $\bar I(x)\!=\!2\gamma_0I(\omega)/\hbar$ and $\bar
I_{0}\!=\!\bar\Gamma_f(\varepsilon_f)/2\pi$. After some algebra,
this results in
\begin{equation}
\bar{I}(x)\approx\frac{\bar{I}_{0}(\varepsilon_f)\,[(x-\varepsilon_f)^{2}+\Delta
x_p^2]}{[(x-\varepsilon_f)^{2}-X_f^{2}/4]^{2}+(x-\varepsilon_f)^{2}(\Delta
x_p^2+\Delta\varepsilon_f^2)}\,, \label{Ixfin}
\end{equation}
where $\Delta x_p^2>\Delta\varepsilon_f^2$. The summation sign
over the exciton internal states is omitted since only one
internal state contributes to the emission intensity in the
vicinity of the sharp plasmon resonance.

The line shape in Eq.~(\ref{Ixfin}) is mainly determined by the
coupling parameter $(X_f/\Delta x_p)^2$. It is clearly seen to be
of a symmetric two-peak structure in the strong coupling regime
where $(X_f/\Delta x_p)^{2}\gg1$. Testing it for extremum, we
obtain the peak frequencies to be
\[
x_{1,2}=\varepsilon_f\pm\frac{X_f}{2}\sqrt{\sqrt{1+
8\left(\!\frac{\Delta
x_p}{X_f}\right)^{2}}\!\!-4\left(\!\frac{\Delta
x_p}{X_f}\right)^{2}}
\]
[terms $\sim\!(\Delta x_p)^2(\Delta\varepsilon_f)^2/X_f^4$ are
neglected], with the Rabi splitting $x_{1}-x_{2}\!\approx\!X_f$.
In the weak coupling regime~where $(X_f/\Delta x_p)^{2}\ll1$, the
frequencies $x_1$ and $x_2$ become complex, indicating that there
are no longer peaks at these frequencies. As this takes place,
Eq.~(\ref{Ixfin}) is approximated with the weak coupling
condition, the fact that $x\!\sim\!\varepsilon_f$, and
$X_f^2=2\Delta x_p\bar\Gamma_f(\varepsilon_f)$, to yield the
Lorentzian
\[
\tilde{I}(x)\approx\frac{\bar{I}_{0}(\varepsilon_f)/[1+(\Delta\varepsilon_f/\Delta
x_p)^2]}{(x-\varepsilon_f)^{2}+\left[\bar\Gamma_f(\varepsilon_f)/2\sqrt{1+(\Delta\varepsilon_f^{\!}/\Delta
x_p)^2}\;\right]^2}
\]
peaked at $x=\varepsilon_f$, whose half-width-at-half-maximum~is
slightly narrower, however, than $\bar\Gamma_f(\varepsilon_f)/2$
it should be if the exciton-plasmon relaxation were the only
relaxation mechanism in the system.~The reason is the competing
phonon scattering takes excitons out of resonance with plasmons,
thus decreasing the exciton-plasmon relaxation rate. We therefore
conclude that~the phonon scattering does not affect the exciton
emission/absorption line shape when the exciton-plasmon coupling
is strong (it facilitates the strong coupling regime to occur,
however, as was noticed above), and it narrows the (Lorentzian)
emission/absorption line when the exciton-plasmon coupling is
weak.

\begin{figure}[t]
\epsfxsize=8.65cm\centering{\epsfbox{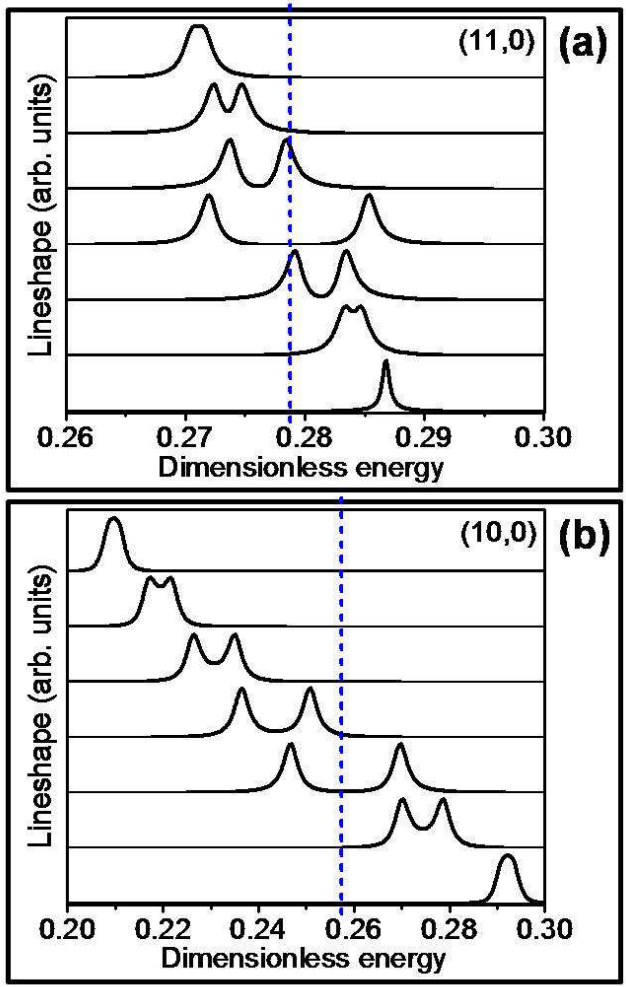}}\caption{(Color
online)~(a),(b)~Exciton absorption/emission lineshapes as the
exciton energies are tuned to the nearest plasmon resonance
energies (vertical dashed lines in here; see Fig.~\ref{fig3} and
left panels in Fig.~\ref{fig4}) in the (11,0) and (10,0)
nano\-tubes, respectively. The dimensionless energy is defined as
[\emph{Energy}]/$2\gamma_0$, according to
Eq.~(\ref{dimless}).}\label{fig7}
\end{figure}

Calculated exciton emission/absorption lineshapes, as given by
Eq.~(\ref{Ixfin}) for the CNs under consideration, are shown in
Fig.~\ref{fig7}~(a) and (b).~The exciton energies are assumed to
be tuned, e.~g. by means of~the quantum confined Stark effect
discussed in Sec.~\ref{sec3}.B, to the nearest plasmon resonances
(shown by the vertical dashed lines in the figure). We used
$\tau_{ph}\!=\!30$~fs as reported in Ref.~\cite{Perebeinos05}. The
line (Rabi) splitting effect is seen to be $\sim0.1$~eV,
indicating the strong exciton-plasmon coupling with the formation
of the mixed surface plasmon-exciton excitations. The splitting is
larger in the smaller diameter nanotubes, and is not masked by the
exciton-phonon scattering.

\section{Conclusions}\label{concl}

We have shown that the strong exciton-surface-plasmon coupling
effect with characteristic exciton absorption line (Rabi)
splitting $\sim\!0.1$~eV exists in small-diameter
($\lesssim\!1$~nm) semiconducting CNs.~The splitting is almost as
large as the typical exciton binding energies in such CNs
($\sim\!0.3-0.8$~eV~\cite{Pedersen03,Pedersen04,Wang05,Capaz}),
and of the same order of magnitude as the exciton-plasmon Rabi
splitting in organic semiconductors
($\sim\!180$~meV~\cite{Bellessa}).~It is much larger than the
exciton-polariton Rabi splitting in semiconductor microcavities
($\sim\!140-400\,\mu\mbox{eV}\,$\cite{Reithmaier,Yoshie,Peter}),
or the exciton-plasmon Rabi splitting in hybrid
semiconductor-metal nanoparticle molecules~\cite{Govorov}.

Since the formation of the strongly coupled mixed exciton-plasmon
excitations is only possible if the exciton total energy is in
resonance with the energy of an interband surface plasmon mode, we
have analyzed possible ways to tune the exciton energy to the
nearest surface plasmon resonance. Specifically, the exciton
energy may be tuned to the nearest plasmon resonance in ways used
for the excitons in semiconductor quantum microcavities
--- thermally (by elevating sample
temperature)~\cite{Reithmaier,Yoshie,Peter}, and/or
electrostatically~\cite{MillerPRL,Miller,Zrenner,Krenner} (via the
quantum confined Stark effect with an external electrostatic field
applied perpendicular to the CN axis).~The two possibilities
influence the different degrees of freedom of the quasi-1D exciton
--- the (longitudinal) kinetic energy and the excitation energy,
respectively.

We have studied how the perpendicular electrostatic field affects
the exciton excitation energy and interband plasmon resonance
energy (the quantum confined Stark effect). Both of them are shown
to shift to the red due to the decrease in the CN band gap as the
field increases. However, the exciton red shift is much less than
the plasmon one because of the decrease in the absolute value of
the negative binding energy, which contributes largely to the
exciton excitation energy. The exciton excitation energy and
interband plasmon energy approach as the field increases, thereby
bringing the total exciton energy in resonance with the plasmon
mode due to the non-zero longitudinal kinetic energy term at
finite temperature.

Lastly, the noteworthy message we would like to deliver in this
paper is that the strong exciton-surface-plasmon coupling we
predict here occurs in an individual CN as opposed to various
artificially fabricated hybrid plasmonic nanostructures mentioned
above. We strongly believe this phenomenon, along with its
tunability feature via the quantum confined Stark effect we have
demonstrated, opens up new paths for the development of CN based
tunable optoelectronic device applications in areas such as
nanophotonics, nanoplasmonics, and cavity QED. One straightforward
application like this is the CN photoluminescence control by means
of the exciton-plasmon coupling tuned electrostatically via the
quantum confined Stark effect.~This complements the microcavity
controlled CN infrared emitter application reported
recently\cite{Avouris08}, offering the advantage of less stringent
fabrication requirements at the same time since the planar
photonic microcavity is no longer required. Electrostatically
controlled coupling of two spatially separated (weakly localized)
excitons to the same nanotube's plasmon resonance would result in
their entanglement~\cite{Bondarev07,Bondarev07jem,Bondarev07os},
the phenomenon that paves the way for CN based solid-state quantum
information applications. Moreover, CNs combine advantages such as
electrical conductivity, chemical stability, and high surface area
that make them excellent potential candidates for a variety of
more practical applications, including efficient solar energy
conversion~\cite{Trancik}, energy storage~\cite{Shimoda}, and
optical nanobiosensorics~\cite{AngChem09}. However, the
photoluminescence quantum yield of individual CNs is relatively
low, and this hinders their uses in the aforementioned
applications. CN bundles and films are proposed to be used to
surpass the poor performance of individual tubes. The theory of
the exciton-plasmon coupling we have developed here, being
extended to include the inter-tube interaction, complements
currently available 'weak-coupling' theories of the
exciton-plasmon interactions in low-dimensional
nanostructures~\cite{Govorov,GovorovPRB} with the very important
case of the strong coupling regime. Such an extended theory
(subject of our future publication) will lay the foundation for
understanding inter-tube energy transfer mechanisms that affect
the efficiency of optoelectronic devices made of CN bundles and
films, as well as it will shed more light on the recent
photoluminescence experiments with CN
bundles~\cite{Munich,Ferrari} and multi-walled CNs~\cite{Hirori},
revealing their potentialities for the development of high-yield,
high-performance optoelectronics applications with CNs.

\acknowledgments

The work is supported by NSF (grants ECS-0631347 and HRD-0833184).
L.M.W. and K.T. acknowledge support from DOE (grant
DE-FG02-06ER46297). Helpful discussions with Mikhail Braun
(St.-Peterburg U., Russia), Jonathan Finley (WSI, TU Munich,
Germany), and Alexander Govorov (Ohio U., USA) are gratefully
acknowledged.

\appendix

\section{~Exciton interaction with the surface electromagnetic field}\label{appA}

We follow our recently developed QED formalism to describe
vacuum-type EM effects in the presence of quasi-1D absorbing and
dispersive
bodies~\cite{Bondarev02,Bondarev04,Bondarev04pla,Bondarev04ssc,Bondarev05,Bondarev06trends}.
The treatment begins with the most general EM interaction of the
surface charge fluctuations with the quantized surface EM field of
a single-walled CN. No external field is assumed to be applied.
The CN is modelled by a neutral, infinitely long, infinitely thin,
anisotropically conducting cylinder. Only the axial conductivity
of the CN, $\sigma_{zz}$,~is taken into account, whereas the
azimuthal one, $\sigma_{\varphi\varphi}$, is neglected being
strongly suppressed by the transverse depolarization
effect~\cite{Benedict,Tasaki,Li,Marinop,Ando,Kozinsky}. Since the
problem has the cylindrical symmetry, the orthonormal cylindrical
basis $\{\mathbf{e}_{r},\mathbf{e}_{\varphi},\mathbf{e}_{z}\}$ is
used with the vector $\mathbf{e}_{z}$ directed along the nanotube
axis as shown in Fig.~\ref{fig1}. The interaction has the
following form (Gaussian system of units)
\begin{eqnarray}
\hat{H}_{int}=\hat{H}_{int}^{(1)}+\hat{H}_{int}^{(2)}\hskip2.3cm\label{Hint12}\\[0.3cm]
=-\sum_{\mathbf{n},i}\frac{q_i}{m_{i}c}\hat{\mathbf{A}}(\mathbf{n}+
\hat{\mathbf{r}}_\mathbf{n}^{(i)})\!\cdot\!\left[\hat{\mathbf{p}}_\mathbf{n}^{(i)}-
\frac{q_i}{2c}\hat{\mathbf{A}}(\mathbf{n}+\hat{\mathbf{r}}_\mathbf{n}^{(i)})\right]\nonumber\\
+\sum_{\mathbf{n},i}q_i\hat{\varphi}(\mathbf{n}+\hat{\mathbf{r}}_\mathbf{n}^{(i)}),\hskip2.3cm\nonumber
\end{eqnarray}
where $c$ is the speed of light, $m_i$, $q_i$,
$\hat{\mathbf{r}}_\mathbf{n}^{(i)}$, and
$\hat{\mathbf{p}}_\mathbf{n}^{(i)}$ are, respectively, the masses,
charges, coordinate operators and momenta operators of the
particles (electrons and nucleus) residing at the lattice site
$\mathbf{n}\!=\!\mathbf{R}_n\!=\!\{R_{CN},\varphi_n,z_n\}$
associated with a carbon atom (see Fig.~\ref{fig1}) on the surface
of the CN of radius $R_{CN}$. The summation is taken over the
lattice sites, and may be substituted with the integration over
the CN surface using Eq.~(\ref{sumrule}). The vector potential
operator $\hat{\mathbf{A}}$ and the scalar potential operator
$\hat{\varphi}$ represent the nanotube's transversely polarized
and longitudinally polarized surface EM modes, respectively. They
are written in the Schr\"{o}dinger picture as follows
\begin{eqnarray}
\hat{\mathbf{A}}(\mathbf{n})=\int_0^\infty\!\!\!\!\!d\omega\frac{c}{i\omega}\,
\hat{\underline{\mathbf{E}}}^\perp(\mathbf{n},\omega)+h.c.,\label{A}\\[0.2cm]
-\bm{\nabla}_{\!\mathbf{n}}\,\hat{\varphi}(\mathbf{n})=\int_0^\infty\!\!\!\!\!d\omega\,
\hat{\underline{\mathbf{E}}}^\parallel(\mathbf{n},\omega)+h.c..\label{phi}
\end{eqnarray}
We use the Coulomb gauge whereby
$\bm{\nabla}_{\!\mathbf{n}}\!\cdot\!\hat{\mathbf{A}}(\mathbf{n})\!=\!0$,
or equivalently,
$[\hat{\mathbf{p}}_\mathbf{n}^{(i)}\!,\hat{\mathbf{A}}(\mathbf{n}+\hat{\mathbf{r}}_{\mathbf{n}}^{(i)})]=0$.

The total electric field operator of the CN-modified~EM field is
given for an arbitrary $\mathbf{r}$ in the Schr\"{o}dinger picture
by
\begin{eqnarray}
\hat{\mathbf{E}}(\mathbf{r})=\!\int_0^\infty\!\!\!\!\!d\omega\,
\hat{\underline{\mathbf{E}}}(\mathbf{r},\omega)+h.c.\hskip0.7cm\label{Er}\\[0.2cm]
=\!\int_0^\infty\!\!\!\!\!d\omega\,[\hat{\underline{\mathbf{E}}}^\perp(\mathbf{r},\omega)+
\hat{\underline{\mathbf{E}}}^\parallel(\mathbf{r},\omega)]+h.c.\nonumber
\end{eqnarray}
with the transversely (longitudinally) polarized Fourier-domain
field components defined as
\begin{equation}
\underline{\hat{\mathbf{E}}}^{\perp(\parallel)}(\mathbf{r},\omega)=
\!\int\!d\mathbf{r}^{\prime}\,\bm{\delta}^{\perp(\parallel)}
(\mathbf{r}-\mathbf{r}^{\prime})\cdot\underline{\hat{\mathbf{E}}}
(\mathbf{r}^{\prime}\!,\omega), \label{elperpar}
\end{equation}
where

\begin{eqnarray}
\delta^{\parallel}_{\alpha\beta}(\mathbf{r})=
-\nabla_{\alpha}\nabla_{\beta}{1\over{4\pi r}}\,,\hskip0.2cm
\label{deltapar}\\[0.2cm]
\delta^{\perp}_{\alpha\beta}(\mathbf{r})=\delta_{\alpha\beta}\,
\delta(\mathbf{r})-\delta^{\parallel}_{\alpha\beta}(\mathbf{r})\nonumber
\end{eqnarray}
are the longitudinal and transverse dyadic $\delta$-functions,
respectively. The total field operator (\ref{Er}) satisfies the
set of the Fourier-domain Maxwell equations
\begin{eqnarray}
\bm{\nabla}\times\underline{\hat{\mathbf{E}}}(\mathbf{r},\omega)=
ik\,\underline{\hat{\mathbf{H}}}(\mathbf{r},\omega),\hskip1.2cm
\label{MaxwelE}\\[0.2cm]
\bm{\nabla}\times\underline{\hat{\mathbf{H}}}(\mathbf{r},\omega)=
-ik\,\underline{\hat{\mathbf{E}}}(\mathbf{r},\omega)+
{4\pi\over{c}}\,\underline{\hat{\mathbf{I}}}(\mathbf{r},\omega),
\label{MaxwelH}
\end{eqnarray}
where $\underline{\hat{\mathbf{H}}}=(ik)^{-1}
\bm{\nabla}\times\underline{\hat{\mathbf{E}}}$ is the magnetic
field operator, $k=\omega/c$, and
\begin{equation}
\underline{\hat{\mathbf{I}}}(\mathbf{r},\omega)=\sum_{\mathbf{n}}
\delta(\mathbf{r}-\mathbf{n})\,\underline{\hat{\mathbf{J}}\!}\,(\mathbf{n},\omega),
\label{Irw}
\end{equation}
is the exterior current operator with the current density defined
as follows
\begin{equation}
\underline{\hat{\mathbf{J}}\!}\,(\mathbf{n},\omega)\!=
\!\sqrt{\hbar\omega\,\mbox{Re}\sigma_{zz}(R_{CN},\omega)\over{\pi}}\,
\hat{f}(\mathbf{n},\omega)\textbf{e}_{z} \label{currentCN}
\end{equation}
to ensure preservation of the fundamental QED equal-time
commutation relations (see, e.g.,~\cite{VogelWelsch}) for the EM
field components in the presence of a CN. Here, $\sigma_{zz}$ is
the CN surface axial conductivity per unit length, and
$\hat{f}(\mathbf{n},\omega)$ along with its counter-part
$\hat{f}^\dag(\mathbf{n},\omega)$ are the scalar bosonic field
operators which annihilate and create, respectively,
single-quantum EM field excitations of frequency $\omega$ at the
lattice site $\mathbf{n}$ of the CN surface. They satisfy the
standard bosonic commutation relations
\begin{eqnarray}
[\hat{f}(\mathbf{n},\omega),\hat{f}^\dag(\mathbf{m},\omega^{\prime})]=
\delta_{\mathbf{n}\mathbf{m}}\,\delta(\omega-\omega^{\prime}),\hskip0.7cm\label{commut}\\[0.3cm]
[\hat{f}(\mathbf{n},\omega),\hat{f}(\mathbf{m},\omega^{\prime})]=
[\hat{f}^\dag(\mathbf{n},\omega),\hat{f}^\dag(\mathbf{m},\omega^{\prime})]=0.\nonumber
\end{eqnarray}

One further obtains from Eqs.~(\ref{MaxwelE})--(\ref{currentCN})
that
\begin{equation}
\underline{\hat{\mathbf{E}}}(\mathbf{r},\omega)=
ik{4\pi\over{c}}\sum_{\mathbf{n}}\mathbf{G}(\mathbf{r},\mathbf{n},\omega)
\!\cdot\!\underline{\hat{\mathbf{J}}\!}\,(\mathbf{n},\omega),
\label{Erw}
\end{equation}
and, according to Eqs.~(\ref{Er}) and (\ref{elperpar}),
\begin{equation}
\underline{\hat{\mathbf{E}}}^{\perp(\parallel)}(\mathbf{r},\omega)=
ik{4\pi\over{c}}\sum_{\mathbf{n}}\,\!^{\perp(\parallel)}\mathbf{G}(\mathbf{r},\mathbf{n},\omega)
\!\cdot\!\underline{\hat{\mathbf{J}}\!}\,(\mathbf{n},\omega),
\label{Erwperpar}
\end{equation}
where $^\perp\mathbf{G}$ and $^\parallel\mathbf{G}$ are the
transverse part and the longitudinal part, respectively, of the
total Green tensor
$\mathbf{G}=\,\!^{\perp}\mathbf{G}+\,\!^{\parallel}\mathbf{G}$ of
the classical EM field in the presence of the CN. This tensor
satisfies the equation
\begin{equation}
\sum_{\alpha=r,\varphi,z}\!\!\!
\left(\bm{\nabla}\!\times\bm{\nabla}\!\times-\,k^{2}\right)_{\!z\alpha}
G_{\alpha z}(\mathbf{r},\mathbf{n},\omega)=
\delta(\mathbf{r}-\mathbf{n}) \label{GreenequCN}
\end{equation}
together with the radiation conditions at infinity and the
boundary conditions on the CN surface.

All the 'discrete' quantities in
Eqs.~(\ref{Irw})--(\ref{GreenequCN}) may be equivalently rewritten
in continuous variables in view of Eq.~(\ref{sumrule}). Being
applied to the identity
$1=\sum_\mathbf{m}\delta_{\mathbf{n}\mathbf{m}}$,
Eq.~(\ref{sumrule}) yields
\begin{equation}
\delta_{\mathbf{n}\mathbf{m}}=S_0\,\delta(\mathbf{R}_n\!-\mathbf{R}_m).
\label{deltacontin}
\end{equation}
This requires to redefine
\begin{equation}
\hat{f}(\mathbf{n},\omega)=\sqrt{S_0}\,\hat{f}(\mathbf{R}_n,\omega),~
\hat{f}^\dag(\mathbf{n},\omega)=\sqrt{S_0}\,\hat{f}^\dag(\mathbf{R}_n,\omega)
\label{fcontin}
\end{equation}
in the commutation relations (\ref{commut}). Similarly, from
Eq.~(\ref{Erw}), in view of Eqs.~(\ref{sumrule}),
(\ref{currentCN}) and (\ref{fcontin}), one obtains
\begin{equation}
\mathbf{G}(\mathbf{r},\mathbf{n},\omega)=\sqrt{S_0}\,\mathbf{G}(\mathbf{r},\mathbf{R}_n,\omega),
\label{Gcontin}
\end{equation}
which is also valid for the transverse and longitudinal Green
tensors in Eq.~(\ref{Erwperpar}).

Next, we make the series expansions of the interactions
$\hat{H}_{int}^{(1)}$ and $\hat{H}_{int}^{(2)}$ in
Eq.~(\ref{Hint12}) about the lattice site $\mathbf{n}$ to the
first non-vanishing terms,
\begin{eqnarray}
\hat{H}_{int}^{(1)}\approx-\sum_{\mathbf{n},i}\frac{q_i}{m_{i}c}\hat{\mathbf{A}}(\mathbf{n})\cdot
\hat{\mathbf{p}}_\mathbf{n}^{(i)}+\sum_{\mathbf{n},i}\frac{q_i^2}{2m_ic^2}\hat{\mathbf{A}}^2(\mathbf{n}),
\hskip0.5cm\label{Hint1app}\\[0.2cm]
\hat{H}_{int}^{(2)}\approx\sum_{\mathbf{n},i}q_i\mathbf{\nabla}_{\!\mathbf{n}}\,
\hat{\varphi}(\mathbf{n})\cdot\hat{\mathbf{r}}_\mathbf{n}^{(i)},\hskip2.0cm\label{Hint2app}
\end{eqnarray}
and introduce the single-lattice-site Hamiltonian
\begin{equation}
\hat{H}_\mathbf{n}=\varepsilon_0|0\rangle\langle0|+
\sum_f(\varepsilon_0+\hbar\omega_f)|f\rangle\langle\,\!f|\label{Hn}
\end{equation}
with the completeness relation
\begin{equation}
|0\rangle\langle0|+\sum_f|f\rangle\langle\,\!f|=\hat{I}.\label{completeness}
\end{equation}
Here, $\varepsilon_0$ is the energy of the ground state
$|0\rangle$ (no exciton excited) of the carbon atom associated
with the lattice site $\mathbf{n}$, $\varepsilon_0+\hbar\omega_f$
is the~energy of the excited carbon atom in the quantum state
$|f\rangle$ with one $f$-internal-state exciton formed of the
energy $E_{exc}^{(f)}=\hbar\omega_f$. In view of Eqs.~(\ref{Hn})
and (\ref{completeness}), one has
\begin{eqnarray}
\hat{\mathbf{p}}_\mathbf{n}^{(i)}=m_{i}\frac{d\,\hat{\mathbf{r}}_\mathbf{n}^{(i)}}{dt}
=\frac{m_{i}}{i\hbar}\,[\hat{\mathbf{r}}_\mathbf{n}^{(i)},\hat{H}_\mathbf{n}]=
\frac{m_{i}}{i\hbar}\,\hat{I}\,[\hat{\mathbf{r}}_\mathbf{n}^{(i)},\hat{H}_\mathbf{n}]\,\hat{I}
\hskip0.5cm\nonumber\\[0.2cm]
\approx\frac{m_{i}}{i\hbar}\sum_f\hbar\omega_f\!
\left(\langle0|\hat{\mathbf{r}}_\mathbf{n}^{(i)}|f\rangle\,\!B_{\mathbf{n},f}\!-
\langle\,\!f|\hat{\mathbf{r}}_\mathbf{n}^{(i)}|0\rangle\,\!B^\dag_{\mathbf{n},f}\right)
\hskip0.7cm\label{pni}
\end{eqnarray}
and
\begin{equation}
\hat{\mathbf{r}}_\mathbf{n}^{(i)}=\hat{I}\,\hat{\mathbf{r}}_\mathbf{n}^{(i)}\hat{I}
\approx\sum_f\!\left(\langle0|\hat{\mathbf{r}}_\mathbf{n}^{(i)}|f\rangle\,\!B_{\mathbf{n},f}
+\langle\,\!f|\hat{\mathbf{r}}_\mathbf{n}^{(i)}|0\rangle\,\!B^\dag_{\mathbf{n},f}\right)\!,\label{rni}
\end{equation}
where
$\langle0|\hat{\mathbf{r}}_\mathbf{n}^{(i)}|f\rangle=\langle\,\!f|\hat{\mathbf{r}}_\mathbf{n}^{(i)}|0\rangle$
in view of the hermitian and real character of the coordinate
operator. The operators $B_{\mathbf{n},f}\!=\!|0\rangle\langle f|$
and $B^\dag_{\mathbf{n},f}\!=\!|f\rangle\langle 0|$ create and
annihilate, respectively, the $f$-internal-state exciton at the
lattice site $\mathbf{n}$, and exciton-to-exciton transitions are
neglected. In addition, we also have
\begin{equation}
\delta_{ij}\delta_{\alpha\beta}={i\over{\hbar}}\,
[(\hat{\mathbf{p}}_\mathbf{n}^{(i)})_{\alpha},(\hat{\mathbf{r}}_\mathbf{n}^{(j)})_{\beta}],
\label{prcommut}
\end{equation}
where $\alpha,\beta=r,\varphi,z$. Substituting these into
Eqs.~(\ref{Hint1app}) and (\ref{Hint2app}) [commutator
(\ref{prcommut}) goes into the second term of Eq.~(\ref{Hint1app})
which is to be pre-transformed~as follows
$\sum_{i,j,\alpha,\beta}q_iq_j\hat{A}(\mathbf{n})_\alpha
\hat{A}(\mathbf{n})_\beta\delta_{ij}\delta_{\alpha\beta}/2m_ic^2$],
one arrives at the following (electric dipole) approximation of
Eq.~(\ref{Hint12})
\begin{eqnarray}
\hat{H}_{int}=\hat{H}_{int}^{(1)}+\hat{H}_{int}^{(2)}\hskip2.5cm\label{Hint12dipole}\\[0.3cm]
=-\sum_{\mathbf{n},f}\frac{i\omega_f}{c}\,\textbf{d}^f_{\mathbf{n}}\cdot\hat{\textbf{A}}(\textbf{n})
\!\left[B^\dag_{\mathbf{n},f}\!-B_{\mathbf{n},f}+\frac{i}{\hbar\,\!c}\,\textbf{d}^f_{\mathbf{n}}\cdot\hat{\textbf{A}}(\textbf{n})
\right]\nonumber\\
+\sum_{\mathbf{n},f}\textbf{d}^f_{\mathbf{n}}\cdot\bm{\nabla}_{\!\mathbf{n}}\,
\hat{\varphi}(\mathbf{n})\left(B^\dag_{\mathbf{n},f}+B_{\mathbf{n},f}\right)\hskip1.5cm\nonumber
\end{eqnarray}
with
$\textbf{d}^f_{\mathbf{n}}=\langle0|\hat{\mathbf{d}}_\mathbf{n}|f\rangle=\langle\,\!f|\hat{\mathbf{d}}_\mathbf{n}|0\rangle$,
where
$\hat{\mathbf{d}}_\mathbf{n}=\sum_iq_i\hat{\mathbf{r}}_\mathbf{n}^{(i)}$~is
the total electric dipole moment operator of the particles
residing at the lattice site~$\mathbf{n}$.

The Hamiltonian~(\ref{Hint12dipole}) is seen to describe the
vacuum-type exciton interaction with the surface EM field (created
by the charge fluctuations on the nanotube surface). The last term
in the square brackets does not depend on the exciton operators,
and therefore results in the constant energy shift which can be
safely neglected. We then arrive, after using Eqs.~(\ref{A}),
(\ref{phi}), (\ref{currentCN}), and (\ref{Erwperpar}), at the
following second quantized interaction Hamiltonian
\begin{eqnarray}
\hat{H}_{int}=\sum_{\mathbf{n},\mathbf{m},f}\int_{0}^{\infty}\!\!\!\!\!d\omega\,[\,
\mbox{g}_f^{(+)}(\mathbf{n},\mathbf{m},\omega)B^\dag_{\mathbf{n},f}\hskip0.6cm\nonumber\\[0.1cm]
-\;\mbox{g}_f^{(-)}(\mathbf{n},\mathbf{m},\omega)B_{\mathbf{n},f}\,]\,
\hat{f}(\mathbf{m},\omega)+h.c.,\label{HintApp}
\end{eqnarray}
where
\begin{equation}
\mbox{g}_f^{(\pm)}(\mathbf{n},\mathbf{m},\omega)=
\mbox{g}_f^{\perp}(\mathbf{n},\mathbf{m},\omega)\pm
{\omega\over{\omega_f}}\,\mbox{g}_f^{\parallel}(\mathbf{n},\mathbf{m},\omega)
\label{gpmApp}
\end{equation}
with
\begin{eqnarray}
\mbox{g}_f^{\perp(\parallel)}(\mathbf{n},\mathbf{m},\omega)=
-i\frac{4\omega_f}{c^{2}}\sqrt{\pi\hbar\omega\,\mbox{Re}\,\sigma_{zz}(R_{CN},\omega)}
\hskip0.5cm\nonumber\\[0.2cm]
\times\!\!\!\sum_{\alpha=r,\varphi,z}\!\!\!(\textbf{d}^f_{\mathbf{n}})_\alpha\,
^{\perp(\parallel)}G_{\alpha\,\!z}(\mathbf{n},\mathbf{m},\omega),\hskip0.8cm\label{gperpparApp}
\end{eqnarray}
and
\begin{equation}
^{\perp(\parallel)}G_{\alpha\,\!z}(\mathbf{n},\mathbf{m},\omega)=
\!\int\!d\mathbf{r}\;\delta_{\alpha\beta}^{\perp(\parallel)}
(\mathbf{n}-\mathbf{r})\;G_{\beta\,\!z}(\mathbf{r},\mathbf{m},\omega).
\label{GreenperpparApp}
\end{equation}
This yields Eqs.~(\ref{Hint})--(\ref{gperppar}) after the strong
transverse depolarization effect in CNs is taken into account
whereby
$\textbf{d}^f_{\mathbf{n}}\approx(\textbf{d}^f_{\mathbf{n}})_z\mathbf{e}_z$.

\section{~Green tensor of the surface electromagnetic field}\label{appB}

Within the model of an infinitely thin, infinitely long,
anisotropically conducting cylinder we utilize here, the classical
EM field Green tensor is found by expanding the solution to the
Green equation (\ref{GreenequCN}) in series in cylindrical
coordinates, and then imposing the appropriately chosen boundary
conditions on the CN surface to determine the Wronskian
normalization constant (see, e.g., Ref.~\cite{Jackson}).

After the EM field is divided into the transversely and
longitudinally polarized components according to
Eqs.~(\ref{Er})--(\ref{deltapar}), the Green equation
(\ref{GreenequCN}) takes the form
\begin{eqnarray}
\sum_{\alpha=r,\varphi,z}\!\!\!
\left(\bm{\nabla}\!\times\bm{\nabla}\!\times-\,k^{2}\right)_{\!z\alpha}&&\mbox{\hskip-0.8cm}
\left[\;^{\!\!\perp}G_{\alpha
z}(\mathbf{r},\mathbf{n},\omega)+\,^{\!\!\parallel}G_{\alpha
z}(\mathbf{r},\mathbf{n},\omega)\right]\nonumber\\
&=&\delta(\mathbf{r}-\mathbf{n}) \label{Greenperparapp}
\end{eqnarray}
with the two additional constraints,
\begin{equation}
\sum_{\alpha=r,\varphi,z}\!\!\!\nabla_{\alpha}\,^{\!\!\perp}G_{\alpha
z}(\mathbf{r},\mathbf{n},\omega)=0 \label{Gperpgaugeapp}
\end{equation}
and
\begin{equation}
\sum_{\beta,\gamma=r,\varphi,z}\!\!\!
\epsilon_{\alpha\beta\gamma}\nabla_{\beta}\;^{\!\parallel}G_{\gamma
z}(\mathbf{r},\mathbf{n},\omega)=0\,, \label{Gpargaugeapp}
\end{equation}
where $\epsilon_{\alpha\beta\gamma}$ is the totally antisymmetric
unit tensor of rank 3.~Equations (\ref{Gperpgaugeapp}) and
(\ref{Gpargaugeapp}) originate from the divergence-less character
(Coulomb gauge) of the transverse EM component and the curl-less
character of the longitudinal EM component, respectively. The
transverse $^{\perp}G_{\alpha\,\!z}$ and longitudinal
$^{\parallel}G_{\alpha\,\!z}$ Green tensor components are defined
by Eq.~(\ref{GreenperpparApp}) which is the corollary of
Eq.~(\ref{elperpar}) using the Eqs.~(\ref{Erw}) and
(\ref{Erwperpar}). Equation~(\ref{Greenperparapp}) is further
rewritten in view of Eqs.~(\ref{Gperpgaugeapp}) and
(\ref{Gpargaugeapp}), to give the following two independent
equations for $^{\perp}G_{zz}$ and $^{\parallel}G_{zz}$ we need
\begin{eqnarray}
\left(\Delta+k^{2}\right)\,^{\!\!\perp}G_{zz}(\mathbf{r},\mathbf{n},\omega)
=-\delta_{zz}^{\perp}(\mathbf{r}-\mathbf{n})\,,\label{Gzzperpapp}\\[0.2cm]
k^{2}\;^{\!\parallel}G_{zz}(\mathbf{r},\mathbf{n},\omega)
=-\delta_{zz}^{\parallel}(\mathbf{r}-\mathbf{n})\hskip0.6cm\label{Gzzparapp}
\end{eqnarray}
with the transverse and longitudinal delta-functions defined by
Eq.~(\ref{deltapar}).

We use the differential representations for the transverse
$^{\perp}G_{zz}$ and longitudinal $^{\parallel}G_{zz}$ Green
functions of the following form [consistent with
Eq.~(\ref{GreenperpparApp})]
\begin{eqnarray}
^{\perp}G_{zz}(\mathbf{r},\mathbf{n},\omega)=\left(
{1\over{k^{2}}}\nabla_{z}\nabla_{z}+1\right)g(\mathbf{r},\mathbf{n},\omega),
\label{gperapp}\\[0.2cm]
^{\parallel}G_{zz}(\mathbf{r},\mathbf{n},\omega)=
-{1\over{k^{2}}}\nabla_{z}\nabla_{z}\,g(\mathbf{r},\mathbf{n},\omega),\hskip0.4cm
\label{gparapp}
\end{eqnarray}
where $g(\mathbf{r},\mathbf{n},\omega)$ is the scalar Green
function of the Helmholtz equation (\ref{Gzzperpapp}), satisfying
the radiation condition at infinity and the finiteness condition
on the axis of the cylinder. Such a function is known to be given
by the following series expansion
\begin{eqnarray}
g(\mathbf{r},\mathbf{n},\omega)=\frac{\sqrt{S_0}}{4\pi}\,
\frac{e^{ik|\mathbf{r}-\mathbf{R}_n|}}{|\mathbf{r}-\mathbf{R}_n|}
=\frac{\sqrt{S_0}}{(2\pi)^2}\!\sum_{p=-\infty}^{\infty}\!\!e^{ip(\varphi-\varphi_n)}
\hskip0.3cm\label{g0expan}\\
\times\!\int_C\!dh\,I_p(vr)K_p(vR_{CN})\,e^{ih(z-z_n)},\;\;\;\;\;r\leq\,\!R_{CN},
\hskip0.3cm\nonumber
\end{eqnarray}
where $I_{p}$ and $K_{p}$ are the modified cylindric Bessel
functions, $v=v(h,\omega)=\sqrt{h^{2}-k^{2}}$, and we used the
property (\ref{Gcontin}) to go from the discrete variable
$\mathbf{n}$ to the corresponding continuous variable. The
integration contour $C$ goes along the real axis of the complex
plane and envelopes the branch points $\pm k$ of the integrand
from below and from above, respectively. For $r\ge R_{CN}$, the
function $g(\mathbf{r},\mathbf{n},\omega)$ is obtained from
Eq.~(\ref{g0expan}) by means of a simple symbol replacement
$I_{p}\leftrightarrow K_{p}$ in the integrand.

The scalar function (\ref{g0expan}) is to be imposed the boundary
conditions on the CN surface. To derive them, we represent the
classical electric and magnetic field components in terms of the
EM field Green tensor as follows
\begin{eqnarray}
\underline{E\!}_{\,\alpha}(\mathbf{r},\omega)=ik\;\!^{\perp}G_{\alpha\,\!z}(\mathbf{r},\mathbf{n},\omega),
\hskip0.7cm\label{Ealpha}\\[0.2cm]
\underline{H\!}_{\,\alpha}(\mathbf{r},\omega)=-{i\over{k}}\!\!\sum_{\beta,\gamma=r,\varphi,z}
\!\!\!\epsilon_{\alpha\beta\gamma}\nabla_{\beta}E_{\gamma}(\mathbf{r},\omega).
\label{Halpha}
\end{eqnarray}
These are valid for $\mathbf{r}\ne\mathbf{n}$ under the
Coulomb-gauge condition. The boundary conditions are then obtained
from the standard requirements that the tangential electric field
components be continuous across the surface, and the tangential
magnetic field components be discontinuous by an amount
proportional to the free surface current density, which we
approximate here by the (strongest) \emph{axial} component,
$\sigma_{zz}(R_{CN},\omega)$, of the nanotube's surface
conductivity. Under this approximation, one has
\begin{eqnarray}
\underline{E\!}_{\,z}|_+-\underline{E\!}_{\,z}|_-=
\underline{E\!}_{\,\varphi}|_+-\underline{E\!}_{\,\varphi}|_-=0,\hskip1cm\label{boundaryEphiCN}\\[0.2cm]
\underline{H\!}_{\,z}|_+-\underline{H\!}_{\,z}|_-=0,\hskip2.3cm\label{boundaryHzCN}\\[0.1cm]
\underline{H\!}_{\,\varphi}|_+-\underline{H\!}_{\,\varphi}|_-
={4\pi\over{c}}\,\sigma_{zz}(\omega){\underline{E\!}_{\,z}}|_{R_{CN}},
\hskip1cm\label{boundaryHphiCN}
\end{eqnarray}
where $\pm$ stand for $r=R_{CN}\pm\varepsilon$ with the positive
infinitesimal $\varepsilon$. In view of Eqs.~(\ref{Ealpha}),
(\ref{Halpha}) and (\ref{gperapp}), the boundary conditions above
result in the following two boundary conditions for the function
(\ref{g0expan})
\begin{eqnarray}
\left.g\right|_+-\left.g\right|_-=0,\hskip3.cm\label{gbound1}\\[0.2cm]
\left.\frac{\partial\,\!g}{\partial\,\!r}\,\right|_+
-\left.\frac{\partial\,\!g}{\partial\,\!r}\,\right|_-=
-\frac{4\pi\,\!i\,\sigma_{zz}(\omega)}{\omega}\left(\!\frac{\partial^2}{\partial\,\!z^2}\!+
\!k^{2}\!\right)g|_{R_{CN}}.\hskip0.5cm\label{gbound2}
\end{eqnarray}

We see that Eq.~(\ref{gbound1}) is satisfied identically.
Eq.~(\ref{gbound2}) yields the Wronskian of modified Bessel
functions on the left,
$W[I_p(x),K_p(x)]\!=\!I_p(x)K_p^\prime(x)\!-\!K_p(x)I_p^\prime(x)\!=\!-1/x$,
which brings us to the equation
\begin{equation}
-\frac{1}{R_{CN}}=\frac{4\pi\,\!i\,\sigma_{zz}(\omega)}{\omega}\;v^2I_p(vR_{CN})K_p(vR_{CN}).
\label{Wronskian}
\end{equation}
This is nothing but the dispersion relation which determines the
radial wave numbers, $h$, of the CN surface EM modes with given
$p$ and $\omega$. Since we are interested here in the EM field
Green tensor on the CN surface [see Eq.~(\ref{gperpparApp})], not
in particular surface EM modes, we substitute
$I_p(vR_{CN})K_p(vR_{CN})$ from Eq.~(\ref{Wronskian}) into
Eq.~(\ref{g0expan}) with $r=R_{CN}$. This allows us to obtain the
scalar Green function of interest with the boundary conditions
(\ref{gbound1}) and (\ref{gbound2}) taken into account. We have
\begin{equation}
g(\mathbf{R},\mathbf{n},\omega)=-\frac{i\omega\sqrt{S_0}\,\delta(\varphi-\varphi_n)}
{8\pi^2\sigma_{zz}(\omega)R_{CN}}\int_C\!dh\frac{e^{ih(z-z_n)}}{k^2-h^2},\label{gCNsurface}
\end{equation}
where $\mathbf{R}=\{R_{CN},\varphi,z\}$ is an arbitrary point of
the cylindrical surface. Using further the residue theorem to
calculate the contour integral, we arrive at the final expression
of the form
\begin{equation}
g(\mathbf{R},\mathbf{n},\omega)=-\frac{c\,\sqrt{S_0}\,\delta(\varphi-\varphi_n)}
{8\pi\sigma_{zz}(\omega)R_{CN}}\;e^{i\omega|z-z_n|/c},
\label{gCNsurfacefinal}
\end{equation}
which yields
\begin{eqnarray}
^{\perp}G_{zz}(\mathbf{R},\mathbf{n},\omega)\equiv0,\hskip1.0cm\label{Gperpsurface}\\[0.2cm]
^{\parallel}G_{zz}(\mathbf{R},\mathbf{n},\omega)=g(\mathbf{R},\mathbf{n},\omega),\hskip0.4cm
\label{Gparsurface}
\end{eqnarray}
in view of Eqs.~(\ref{gperapp}) and (\ref{gparapp}).

The fact that the transverse Green function (\ref{Gperpsurface})
identically equals zero on the CN surface is related to the
absence of the skin layer in the model of the infinitely thin
cylinder (see, e.g.,~Ref.~\cite{Jackson}). In this model, the
transverse Green function is only non-zero in the near-surface
area where the exciton wave function goes to zero. Thus, only
longitudinally polarized EM modes with the Green
function~(\ref{Gparsurface}) contribute to the exciton surface EM
field interaction on the nanotube surface.

\section{~Diagonalization of the Hamiltonian~(\ref{Htot})--(\ref{gfpar})}\label{appC}

We start with the transformation of the total
Hamiltonian~(\ref{Htot})--(\ref{gfpar}) to the
$\mathbf{k}$-representation using Eqs.~(\ref{Bkf}) and
(\ref{orthog}). The unperturbed part presents no difficulties.
Special care should be given to the interaction matrix element
$\mbox{g}_f^{(\pm)}(\mathbf{n},\mathbf{m},\omega)$ in
Eq.~(\ref{gfpar}). In view of Eqs.~(\ref{Gparsurface}),
(\ref{gCNsurfacefinal}) and (\ref{sumrule}), one has explicitly
\begin{eqnarray}
\mbox{g}_f^{(\pm)}(\mathbf{k},\mathbf{k}^\prime,\omega)=\frac{1}{N}
\sum_{\mathbf{n},\mathbf{m}}\mbox{g}_f^{(\pm)}(\mathbf{n},\mathbf{m},\omega)\,
e^{-i\mathbf{k}\cdot\mathbf{n}+i\mathbf{k}^\prime\cdot\mathbf{m}}\hskip0.5cm\label{gfpmk}\\[0.1cm]
=\pm\frac{i\omega\sqrt{\pi\hbar\omega\,\mbox{Re}\,\sigma_{zz}(\omega)}}
{2\pi\,\!c\,\sigma_{zz}(\omega)R_{CN}}\,\frac{d_z^f}{N}\sqrt{S_0}\hskip1.8cm\nonumber\\[0.1cm]
\times\frac{R_{CN}^2}{NS_0^2}\int_0^{2\pi}\!\!\!\!\!\!\!d\varphi_nd\varphi_m\delta(\varphi_n\!-\varphi_m)\,
e^{-ik_\varphi\varphi_n+ik_\varphi^\prime\varphi_m}\hskip.8cm\nonumber\\[0.1cm]
\times\int_{-\infty}^\infty\!\!\!\!\!\!\!dz_ndz_m\,e^{i\omega|z_n-z_m|/c-ik_zz_n+ik_z^\prime\,\!z_m},\hskip1.4cm\nonumber
\end{eqnarray}
where we have also taken into account the fact that the dipole
matrix element
$(\textbf{d}^f_{\mathbf{n}})_z\!=\!\langle0|(\hat{\mathbf{d}}_\mathbf{n})_z|f\rangle$
is actually the same for all the lattice sites on the CN surface
in view of their equivalence. As a consequence,
$(\textbf{d}^f_{\mathbf{n}})_z=d_z^f/N$ with
$d_z^f=\sum_\mathbf{n}\langle0|(\hat{\mathbf{d}}_\mathbf{n})_z|f\rangle$.

The integral over $\varphi$ in Eq.~(\ref{gfpmk}) is taken in a
standard way to yield
\begin{equation}
\int_0^{2\pi}\!\!\!\!\!\!\!d\varphi_nd\varphi_m\delta(\varphi_n\!-\varphi_m)\,
e^{-ik_\varphi\varphi_n+ik_\varphi^\prime\varphi_m}=2\pi\delta_{k_\varphi\,\!k_\varphi^\prime}.
\label{intphi}
\end{equation}
The integration over $z$ is performed by first writing the
integral in the form
\[
\int_{-\infty}^\infty\!\!\!\!\!\!\!dz_ndz_m...=
\lim_{L\rightarrow\infty}\int_{-L/2}^{L/2}\!\!\!\!\!\!\!dz_n\int_{-L/2}^{L/2}\!\!\!\!\!\!\!dz_m...
\]
($L$ being the CN length), then dividing it into two parts by
means of the equation
\begin{eqnarray}
e^{i\omega|z_n-z_m|/c}=\theta(z_n\!-z_m)\,e^{i\omega(z_n-z_m)/c}\hskip0.3cm\nonumber\\[0.2cm]
+\,\theta(z_m\!-z_n)\,e^{-i\omega(z_n-z_m)/c},\hskip1.cm\nonumber
\end{eqnarray}
and finally by taking simple exponential integrals with allowance
made for the formula
\[
\delta_{k_zk_z^\prime}=\lim_{L\rightarrow\infty}\!\frac{2\sin[L(k_z\!-k_z^\prime)/2]}{L(k_z\!-k_z^\prime)}\,.
\]
After some simple algebra we obtain the result
\begin{eqnarray}
\int_{-\infty}^\infty\!\!\!\!\!\!\!dz_ndz_m\,e^{i\omega|z_n-z_m|/c-ik_zz_n+ik_z^\prime\,\!z_m}
\hskip0.5cm\label{intz}\\
=\lim_{L\rightarrow\infty}\!L^2\!\left\{\!1-\frac{2i\omega/c}{L\,[k_z^2\!-(\omega/c)^2]}\right\}
\delta_{k_zk_z^\prime}.\nonumber
\end{eqnarray}

In view of Eqs.~(\ref{intphi}) and (\ref{intz}), the function
(\ref{gfpmk}) takes the form
\begin{eqnarray}
\mbox{g}_f^{(\pm)}(\mathbf{k},\mathbf{k}^\prime,\omega)=
\pm\frac{i\omega\,d_z^f\sqrt{\pi\,\!S_0\hbar\omega\,\mbox{Re}\,\sigma_{zz}(\omega)}}
{(2\pi)^2c\,\sigma_{zz}(\omega)R_{CN}}\hskip0.5cm\label{gfpmk1}\\[0.2cm]
\times\lim_{L\rightarrow\infty}\!\left\{\!1-\frac{2i\omega/c}{L\,[k_z^2\!-(\omega/c)^2]}\right\}
\delta_{\mathbf{k}\mathbf{k}^\prime}.\hskip1.2cm\nonumber
\end{eqnarray}
We have taken into account here that
$\delta_{k_\varphi\,\!k_\varphi^\prime}\delta_{k_zk_z^\prime}=\delta_{\mathbf{k}\mathbf{k}^\prime}$,
as well as the fact that $(R_{CN}L/NS_0)^2=1/(2\pi)^2$. This can
be further simplified by noticing that only absolute value squared
of the interaction matrix element matters in calculations of
observables. We then have
\[
\left|1-\frac{2i\omega/c}{L\,[k_z^2\!-(\omega/c)^2]}\,\right|^2\!=1+\frac{\alpha}{u^2}
\approx1+\frac{\alpha}{u^2+\alpha^2}
\]
with $u=(ck_z/\omega)^2\!-1$, and $\alpha=(2c/L\omega)^2$ being
the small parameter which tends to zero as $L\rightarrow\infty$.
Using further the formula (see, e.g., Ref.~\cite{Davydov})
\[
\delta(u)=\frac{1}{\pi}\lim_{\alpha\rightarrow0}\frac{\alpha}{u^2+\alpha^2}\;,
\]
and the basic properties of the $\delta$-function, we arrive at
\begin{eqnarray}
\lim_{L\rightarrow\infty}\left|1-\frac{2i\omega/c}{L\,[k_z^2\!-(\omega/c)^2]}\,\right|^2\!=
1+\frac{\pi\,\!c|k_z|}{2}\hskip0.3cm\label{LimitL}\\[0.2cm]
\times\left[\,\delta(\omega+ck_z)+\;\delta(\omega-ck_z)\right].\hskip1.cm\nonumber
\end{eqnarray}
We also have
\begin{equation}
\left|\frac{\sqrt{\mbox{Re}\,\sigma_{zz}(\omega)}}{\sigma_{zz}(\omega)}\,\right|^2\!=
\mbox{Re}\frac{1}{\sigma_{zz}(\omega)}\,.\label{Resigoversig}
\end{equation}
Equation~(\ref{gfpmk1}), in view of Eqs.~(\ref{LimitL}) and
(\ref{Resigoversig}), is rewritten effectively as follows
\begin{eqnarray}
\mbox{g}_f^{(\pm)}(\mathbf{k},\mathbf{k}^\prime,\omega)=
\pm\,\!iD_f(\omega)\,\delta_{\mathbf{k}\mathbf{k}^\prime}
\label{gfpmksqrootform}
\end{eqnarray}
with
\begin{eqnarray}
D_f(\omega)=\frac{\omega\,d_z^f\sqrt{\pi\,\!S_0\hbar\omega\,\mbox{Re}[1/\sigma_{zz}(\omega)]}}
{(2\pi)^2c\,R_{CN}}\hskip1cm\label{Dfomega}\\[0.1cm]
\times\sqrt{1+\frac{\pi\,\!c|k_z|}{2}\left[\,\delta(\omega+ck_z)+\;\delta(\omega-ck_z)\right]}\;.
\hskip0.5cm\nonumber
\end{eqnarray}

In terms of the simplified interaction matrix element
(\ref{gfpmksqrootform}), the $\mathbf{k}$-representation of the
Hamiltonian~(\ref{Htot})--(\ref{gfpar}) takes the following
(symmetrized) form
\begin{equation}
\hat{H}=\frac{1}{2}\sum_\mathbf{k}\hat{H}_\mathbf{k},\label{Htotk}
\end{equation}
where
\begin{eqnarray}
\hat{H}_\mathbf{k}=\sum_fE_f(\mathbf{k})\left(B^\dag_{\mathbf{k},f}B_{\mathbf{k},f}+
B^\dag_{\mathbf{-k},f}B_{\mathbf{-k},f}\right)\hskip0.6cm\label{Htotalk}\\
+\int_0^\infty\!\!\!\!\!d\omega\,\hbar\omega\left[
\hat{f}^\dag(\mathbf{k},\omega)\hat{f}(\mathbf{k},\omega)+
\hat{f}^\dag(-\mathbf{k},\omega)\hat{f}(-\mathbf{k},\omega)\right]\nonumber\\[0.1cm]
+\sum_f\int_0^\infty\!\!\!\!\!d\omega\;iD_f(\omega)\left(
B^\dag_{\mathbf{k},f}+B_{-\mathbf{k},f}\right)\hskip1.2cm\nonumber\\
\times\left[\hat{f}(\mathbf{k},\omega)-\hat{f}^\dag(-\mathbf{k},\omega)\right]+h.c.\hskip1.5cm\nonumber
\end{eqnarray}
with $D_f(\omega)$ given by Eq.~(\ref{Dfomega}). To diagonalize
this Hamiltonian, we follow Bogoliubov's canonical transformation
technique (see, e.g., Ref.~\cite{Davydov}). The canonical
transformation of the exciton and photon operators is of the form
\begin{eqnarray}
B_{\mathbf{k},f}=\!\!\sum_{\mu=1,2}\left[u_\mu(\mathbf{k},\omega_f)\hat{\xi}_\mu(\mathbf{k})
+v_\mu(\mathbf{k},\omega_f)\hat{\xi}^\dag_\mu(-\mathbf{k})\right],\hskip0.5cm\label{Btransf}\\
\hat{f}(\mathbf{k},\omega)=\!\!\sum_{\mu=1,2}\left[u^\ast_\mu(\mathbf{k},\omega)\hat{\xi}_\mu(\mathbf{k})
+v^\ast_\mu(\mathbf{k},\omega)\hat{\xi}^\dag_\mu(-\mathbf{k})\right],\hskip0.5cm\label{ftransf}
\end{eqnarray}
where the new operators, $\hat{\xi}_\mu(\mathbf{k})$ and
$\hat{\xi}^\dag_\mu(\mathbf{k})\!=\![\hat{\xi}_\mu(\mathbf{k})]^\dag$,
annihilate and create, respectively, the coupled exciton-photon
excitations of branch $\mu$ on the nanotube surface. They satisfy
the bosonic commutation relations of the form
\begin{equation}
\left[\hat{\xi}_\mu(\mathbf{k}),\,\hat{\xi}^\dag_{\mu^\prime}(\mathbf{k}^\prime)\right]=
\delta_{\mu\mu^\prime}\delta_{\mathbf{k}\mathbf{k}^\prime},
\label{xikcommut}
\end{equation}
which, along with the reversibility requirement of
Eqs.~(\ref{Btransf}) and (\ref{ftransf}), impose the following
constraints on the transformation functions $u_\mu$ and $v_\mu$
\begin{eqnarray}
\sum_f\left[u^\ast_\mu(\mathbf{k},\omega_f)u_{\mu^\prime}(\mathbf{k},\omega_f)
-v_\mu(\mathbf{k},\omega_f)v^\ast_{\mu^\prime}(\mathbf{k},\omega_f)\right]\hskip0.7cm\nonumber\\
+\int_0^\infty\!\!\!d\omega\left[u_\mu(\mathbf{k},\omega)u^\ast_{\mu^\prime}(\mathbf{k},\omega)
-v^\ast_\mu(\mathbf{k},\omega)v_{\mu^\prime}(\mathbf{k},\omega)\right]=\delta_{\mu\mu^\prime},
\nonumber\\[0.2cm]
\sum_\mu\left[u^\ast_\mu(\mathbf{k},\omega_f)u_\mu(\mathbf{k},\omega_{f^\prime})-
v^\ast_\mu(\mathbf{k},\omega_f)v_\mu(\mathbf{k},\omega_{f^\prime})\right]=\delta_{ff^\prime},
\nonumber\\[0.1cm]
\sum_\mu\left[u^\ast_\mu(\mathbf{k},\omega)u_\mu(\mathbf{k},\omega^\prime)-
v^\ast_\mu(\mathbf{k},\omega)v_\mu(\mathbf{k},\omega^\prime)\right]=\delta(\omega-\omega^\prime).\nonumber
\end{eqnarray}
Here, the first equation guarantees the fulfilment of the
commutation relations (\ref{xikcommut}), whereas the second and
the third ensure that Eqs.~(\ref{Btransf}) and (\ref{ftransf}) are
inverted to yield $\hat{\xi}_\mu(\mathbf{k})$ as given by
Eq.~(\ref{xik}). Other possible combinations of the transformation
functions are identically equal to zero.

The proper transformation functions that diagonalize the
Hamiltonian (\ref{Htotalk}) to bring it to the form
(\ref{Htotdiag}), are determined by the identity
\begin{equation}
\hbar\omega_\mu(\mathbf{k})\,\hat{\xi}_\mu(\mathbf{k})=
\left[\hat{\xi}_\mu(\mathbf{k}),\,\hat{H}_\mathbf{k}\right].
\label{identity}
\end{equation}
Putting Eqs.~(\ref{xik}) and (\ref{Htotalk}) into
Eq.~(\ref{identity}) and using the bosonic commutation relations
for the exciton and photon operators on the right, one obtains
($\mathbf{k}$-argument is omitted for brevity)
\begin{eqnarray}
\left(\hbar\omega_\mu-E_f\right)u^\ast_\mu(\omega_f)=
-i\!\int_0^\infty\!\!\!\!\!d\omega\,D_f(\omega)\left[
u_\mu(\omega)-v^\ast_\mu(\omega)\right],\nonumber\\
\left(\hbar\omega_\mu+E_f\right)v_\mu(\omega_f)=
i\!\int_0^\infty\!\!\!\!\!d\omega\,D_f(\omega)\left[
u_\mu(\omega)-v^\ast_\mu(\omega)\right],\hskip0.1cm\nonumber\\
\hbar\left(\omega_\mu-\omega\right)u_\mu(\omega)=i\!\sum_fD_f(\omega)
\left[u^\ast_\mu(\omega_f)+v_\mu(\omega_f)\right],\hskip0.2cm\nonumber\\
\hbar\left(\omega_\mu+\omega\right)v^\ast_\mu(\omega)=i\!\sum_fD_f(\omega)
\left[u^\ast_\mu(\omega_f)+v_\mu(\omega_f)\right].\hskip0.3cm\nonumber
\end{eqnarray}
These simultaneous equations define the complex transformation
functions $u_\mu$ and $v_\mu$ uniquely. They also define the
dispersion relation (the energies $\hbar\omega_\mu,\;\mu=1,2$) of
the coupled exciton-photon (or exciton-plasmon, to be exact)
excitations on the nanotube surface. Substituting $u_\mu$ and
$v^\ast_\mu$ from the third and forth equations into the first
one, one has
\[
\left[\hbar\omega_\mu-E_f-\frac{4E_f}{\hbar\omega_\mu+E_f}\int_0^\infty\!\!\!\!\!d\omega\,
\frac{\omega|D_f(\omega)|^2}{\hbar(\omega^2_\mu-\omega^2)}\right]u^\ast_\mu(\omega_f)=0,
\]
whereby, since the functions $u^\ast_\mu$ are non-zero, the
dispersion relation we are interested in becomes
\begin{equation}
(\hbar\omega_\mu)^2-E^2_f-4E_f\int_0^\infty\!\!\!\!\!d\omega\,
\frac{\omega|D_f(\omega)|^2}{\hbar(\omega^2_\mu-\omega^2)}=0\,.
\label{disp1}
\end{equation}
The energy $E_0$ of the ground state of the coupled
exciton-plasmon excitations is found by plugging Eq.~(\ref{xik})
into Eq.~(\ref{Htotdiag}) and comparing the result with
Eqs.~(\ref{Htotk}) and (\ref{Htotalk}). This yields
\[
E_0=-\!\!\!\!\!\sum_{\mathbf{k},\,\mu=1,2}\!\!\!\!\hbar\omega_\mu(\mathbf{k})\!
\left[\sum_f|v_\mu(\mathbf{k},\omega_f)|^2+
\!\int_0^\infty\!\!\!\!\!d\omega\,|v_\mu(\mathbf{k},\omega)|^2\right]\!.
\]

Using further $D_f(\omega)$ as explicitly given by
Eq.~(\ref{Dfomega}), the dispersion relation (\ref{disp1}) is
rewritten as follows
\begin{eqnarray}
(\hbar\omega_\mu)^2-E_f^2=\frac{E_fS_0\,|d_z^f|^2}{4\pi^3c^2R^2_{CN}}
\left\{\int_0^\infty\!\!\!\!d\omega\frac{\omega^4\mbox{Re}[1/\sigma_{zz}(\omega)]}
{\omega^2_\mu-\omega^2}\right.\nonumber\\[0.2cm]
\left.+\frac{\pi\,\!(c|k_z|)^5\mbox{Re}[1/\sigma_{zz}(c|k_z|)]}
{\omega^2_\mu-(c|k_z|)^2}\right\}.\hskip1.2cm\nonumber
\end{eqnarray}
Here we have taken into account the general property
$\sigma_{zz}(\omega)=\sigma_{zz}^\ast(-\omega)$, which originates
from the time-reversal symmetry requirement, in the second term on
the right hand side. This term comes from the two delta functions
in $|D_f(\omega)|^2$, and describes the contribution of the
spatial dispersion (wave-vector dependence) to the formation of
the exciton-plasmons. We neglect this term in what follows because
the spatial dispersion is neglected in the nanotube's axial
surface conductivity in our model, and, secondly, because it is
seen to be very small for not too large excitonic wave vectors.
Thus, converting to the dimensionless variables (\ref{dimless}),
we arrive at the dispersion relation (\ref{dispeq}) with the
exciton spontaneous decay (recombination) rate and the plasmon DOS
given by Eqs.~(\ref{Gamma0f}) and (\ref{plDOS}), respectively.

Lastly, bearing in mind that the delta functions in
$|D_f(\omega)|^2$ are responsible for the spatial dispersion which
we neglect in our model, and therefore dropping them out from the
squared interaction matrix element (\ref{gfpmksqrootform}), we
arrive at the property (\ref{gpkk}).

\section{~Effective longitudinal potential in the presence of the perpendicular electrostatic field}\label{appD}

Here we analyze the set of equations (\ref{wfe})--(\ref{wfexc}),
and show that the attractive cusp-type cutoff potential
(\ref{Vcutoff}) with the field dependent cutoff parameter
(\ref{z0}) is a uniformly valid approximation for the effective
electron-hole Coulomb interaction potential (\ref{Veff}) in the
exciton binding energy equation (\ref{wfexc}).

We rewrite Eqs.~(\ref{wfe}) and (\ref{wfh}) in the form of a
single equation as follows
\begin{equation}
\left(\frac{d^2}{d\varphi^2}+q^2+p\,\cos\varphi\right)\psi(\varphi)=0\,.
\label{subbandeq}
\end{equation}
Here, $\varphi=\varphi_{e,h}$, $\psi=\psi_{e,h}$,
$q=R_{CN}\sqrt{2m_{e,h}\varepsilon_{e,h}}/\hbar$,~and
$p=\pm2em_{e,h}R^3_{CN}F/\hbar^2$ with the (+)-sign to be taken
for the electron and the (--)-sign to be taken for the hole. We
are interested in the solutions to Eq.~(\ref{subbandeq}) which
satisfy the $2\pi$-periodicity condition
$\psi(\varphi)=\psi(\varphi+2\pi)$. The change of variable
$\varphi=2t$ transfers this equation to the well known Mathieu's
equation (see, e.g., Refs.~\cite{Bateman,Abramovitz}), reducing
the solution's period by the factor of two. The exact solutions of
interest are, therefore, given by the odd Mathieu functions
$se_{2m+2}(t=\varphi/2)$ with the eigen values $b_{2m+2}$, where
$m$ is a nonnegative integer (notations of Ref.~\cite{Bateman}).
These are the solutions to the Sturm-Liouville problem with
boundary conditions on functions, not on their derivatives.

It is easier to estimate the $z$-dependence of the potential
(\ref{Veff}) if the functions $\psi_{e,h}(\varphi_{e,h})$ are
known explicitly. So, we do solve Eq.~(\ref{subbandeq}) using the
second order perturbation theory in the external field (the term
$p\cos\varphi$).~The second order field corrections are also of
practical importance in the most of experimental applications.

The unperturbed problem yields the two linearly independent
normalized eigen functions and the eigen values as follows
\begin{equation}
\psi_j^{(0)}(\varphi)=\frac{\exp(\pm\,\!ij\varphi)}{\sqrt{2\pi}}\,,\;\;\;
q=j=\frac{R_{CN}}{\hbar}\sqrt{2m_{e,h}\varepsilon^{(0)}_{e,h}}
\label{unperturbed}
\end{equation}
with $j$ being a nonnegative integer. The energies
$\varepsilon^{(0)}_{e,h}(j)$ are doubly degenerate with the
exception of $\varepsilon^{(0)}_{e,h}(0)=0$, which we will discard
since it results in the zero unperturbed band gap according to
Eq.~(\ref{Egkfi}).~The perturbation $p\cos\varphi$ does not lift
the degeneracy of the unperturbed states. Therefore, we use the
standard nondegenerate perturbation theory with the basis wave
functions set above (plus sign selected for definiteness) to
calculate the energies and the wave functions to the second order
in perturbation. The standard procedure (see,
e.g.,~Ref.~\cite{LandauQM}) yields
\begin{widetext}
\begin{eqnarray}
\psi_{j\,e,h}(\varphi_{e,h})=\left(1-\left\{\frac{\vartheta(j-2)}{\left[(j-1)^2-j^2\right]^2}+
\frac{1}{\left[(j+1)^2-j^2\right]^2}\right\}\frac{m^2_{e,h}e^2R_{CN}^6}{2\hbar^4}F^2\right)
\psi^{(0)}_{j\,e,h}(\varphi_{e,h})\label{psiehfield}\\[0.1cm]
\pm\left[\frac{\vartheta(j-2)\psi^{(0)}_{j-1\,e,h}(\varphi_{e,h})}{(j-1)^2-j^2}+
\frac{\psi^{(0)}_{j+1\,e,h}(\varphi_{e,h})}{(j+1)^2-j^2}\right]\frac{m_{e,h}eR_{CN}^3}{\hbar^2}F\hskip2.3cm\nonumber\\[0.1cm]
+\left\{\frac{\vartheta(j-2)\vartheta(j-3)\psi^{(0)}_{j-2\,e,h}(\varphi_{e,h})}{[(j-1)^2-j^2][(j-2)^2-j^2]}+
\frac{\psi^{(0)}_{j+2\,e,h}(\varphi_{e,h})}{[(j+1)^2-j^2][(j+2)^2-j^2]}\right\}
\frac{m^2_{e,h}e^2R_{CN}^6}{\hbar^4}F^2\,.\nonumber
\end{eqnarray}
\end{widetext}
Here, $j$ is a positive integer, and the theta-functions ensure
that $j=1$ is the ground state of the system. The corresponding
energies are as follows
\begin{equation}
\varepsilon_{e,h}=\frac{\hbar^2j^2}{2m_{e,h}R_{CN}^2}-\frac{m_{e,h}e^2R_{CN}^4w_j}{2\hbar^2}F^2
\end{equation}
with $w_j$ given by Eq.~(\ref{DeltaF}), thus, according to
Eq.~(\ref{Egkfi}), resulting in the nanotube's band gap as given
by Eq.~(\ref{EgF}).

From Eq.~(\ref{psiehfield}), in view of Eq.~(\ref{unperturbed}),
we have the following to the second order in the field
\begin{eqnarray}
|\psi_e(\varphi_e)|^2|\psi_h(\varphi_h)|^2\approx\frac{1}{4\pi^2}\hskip1.5cm\label{psie2psih2}\\[0.1cm]
\times\left[1-2\left(m_h\cos\varphi_h-m_e\cos\varphi_e\right)\frac{eR_{CN}^3w_j}{\hbar^2}F\right.\nonumber\\[0.1cm]
+2\left(m^2_h\cos2\varphi_h+m^2_e\cos2\varphi_e\right)\frac{e^2R_{CN}^6v_j}{\hbar^4}F^2\nonumber\\[0.1cm]
\left.-\,4\mu\,\!M_{ex}\cos\varphi_e\cos\varphi_h\frac{e^2R_{CN}^6w^2_j}{\hbar^4}F^2\right],\hskip0.5cm\nonumber
\end{eqnarray}
where
\begin{eqnarray}
v_j=\!\frac{\vartheta(j-2)}{(j-1)^2\!-j^2}\left\{
\!\frac{\vartheta(j-3)}{(j-2)^2\!-j^2}+\!\frac{1}{(j+1)^2\!-j^2}\right\}\nonumber\\[0.1cm]
+\frac{1}{[(j+1)^2-j^2][(j+2)^2-j^2]}\,.\hskip1.5cm\nonumber
\end{eqnarray}
Plugging Eqs.~(\ref{psie2psih2}) and (\ref{V}) into
Eq.~(\ref{Veff}) and noticing that the integrals involving linear
combinations of the cosine-functions are strongly suppressed due
to the integration over the cosine period, and are therefore
negligible compared to the one involving the quadratic
cosine-combination, we obtain
\begin{eqnarray}
V_{\mbox{\small{eff}}}(z)=-\frac{e^2}{4\pi^2\epsilon}\hskip2.5cm\label{Veffnew}\\[0.1cm]
\times\int_0^{2\pi}\!\!\!\!\!\!d\varphi_e\!\!\int_0^{2\pi}\!\!\!\!\!\!
d\varphi_h\frac{1-2\cos\varphi_e\cos\varphi_h\Delta_j(F)}
{\{z^2+4R_{CN}^2\sin^2[(\varphi_{e}\!-\varphi_{h})/2]\}^{1/2}}\nonumber
\end{eqnarray}
with $\Delta_j(F)$ given by Eq.~(\ref{DeltaF}).

The next step is to perform the double integration in
Eq.~(\ref{Veffnew}). We have to evaluate the two double integrals.
They are
\begin{equation}
I_1=\int_0^{2\pi}\!\!\!\!\!\!d\varphi_e\!\!\int_0^{2\pi}\!\!\!\!\!\!\frac{d\varphi_h}
{\{z^2+4R_{CN}^2\sin^2[(\varphi_{e}\!-\varphi_{h})/2]\}^{1/2}}
\label{I1}
\end{equation}
and
\begin{equation}
I_2=\int_0^{2\pi}\!\!\!\!\!\!d\varphi_e\!\!\int_0^{2\pi}\!\!\!\!\!\!\frac{d\varphi_h\cos\varphi_e\cos\varphi_h}
{\{z^2+4R_{CN}^2\sin^2[(\varphi_{e}\!-\varphi_{h})/2]\}^{1/2}}\;.
\label{I2}
\end{equation}
We first notice that both $I_1$ and $I_2$ can be equivalently
rewritten as follows
\begin{equation}
\int_0^{2\pi}\!\!\!\!\!\!d\varphi_e\!\!\int_0^{2\pi}\!\!\!\!\!\!d\varphi_h...=
2\!\int_0^{2\pi}\!\!\!\!\!\!d\varphi_e\!\!\int_0^{\varphi_e}\!\!\!\!\!\!d\varphi_h...
\label{property}
\end{equation}
due to the symmetry of the integrands with respect to the
$(\varphi_e\!=\!\varphi_h)$-line. Using this property, we
substitute $\varphi_h$ with the new variable
$t=\sin[(\varphi_e-\varphi_h)/2]$ in Eqs.~(\ref{I1}) and
(\ref{I2}). This, after simplifications, yields
\begin{equation}
I_1=4\!\int_0^{2\pi}\!\!\!\!\!\!d\varphi_e\!\int_0^{\sin(\varphi_e/2)}\!\!\!\!\!\!\!\!\!\!\!\!\!\!\!\!\!
\frac{dt}{[(1-t^2)(z^2+4R_{CN}^2t^2)]^{1/2}}\label{I1next}
\end{equation}
and
\begin{equation}
I_2=4\!\int_0^{2\pi}\!\!\!\!\!\!d\varphi_e\cos^2\varphi_e\!\int_0^{\sin(\varphi_e/2)}
\!\!\!\!\!\!\!\!\!\!\!\!\!\!\!\!\!\frac{dt\,(1-2t^2)}{[(1-t^2)(z^2+4R_{CN}^2t^2)]^{1/2}}\;.
\label{I2next}
\end{equation}
Here, the inner integrals are reduced to the incomplete elliptical
integrals of the first and second kinds (see, e.g.,
Ref.~\cite{Abramovitz}).

We continue the evaluation of Eqs.~(\ref{I1next}) and
(\ref{I2next}) by expanding the denominators of the integrands in
series at large and small $|z|$ as compared to the CN diameter
$2R_{CN}$. One has
\begin{eqnarray}
\frac{1}{(z^2+4R_{CN}^2t^2)^{1/2}}\approx\frac{1}{|z|}\left[
1-\frac{1}{2}\left(\frac{2R_{CN}t}{|z|}\right)^2\right.\nonumber\\[0.2cm]
+\left.\frac{3}{8}\left(\frac{2R_{CN}t}{|z|}\right)^4
-\frac{5}{16}\left(\frac{2R_{CN}t}{|z|}\right)^6+...\right]\nonumber
\end{eqnarray}
for $|z|/2R_{CN}\gg1$, and
\begin{eqnarray}
\int_0^{\sin(\varphi_e/2)}
\!\!\!\!\!\!\!\!\!\!\!\!\!\!\!\!\!\frac{dt\,f(t)}{[(1-t^2)(z^2+4R_{CN}^2t^2)]^{1/2}}
=\frac{1}{2R_{CN}}\nonumber\\[0.2cm]\times\lim_{\left(|z|/2R_{CN}\right)\rightarrow0}\;
\int_{|z|/2R_{CN}}^{\sin(\varphi_e/2)}\!\!\!\!\!\!\!dt\frac{f(t)}{t\sqrt{1-t^2}}\hskip0.5cm\nonumber
\end{eqnarray}
for $|z|/2R_{CN}\ll1$ [$f(t)$ is a polynomial function]. Using
these in Eqs.~(\ref{I1next}) and (\ref{I2next}), we arrive at
\begin{eqnarray}
I_1\approx\left\{\!\!\!\begin{array}{l}\frac{\displaystyle4\pi}{\displaystyle\,\!R_{CN}}\!\left[\,
\ln\!\left(\!\frac{\displaystyle4R_{CN}}{\displaystyle|z|}\right)\!
-\frac{\displaystyle1}{\displaystyle4}\left(\!\frac{\displaystyle|z|}{\displaystyle2R_{CN}}\right)^{\!2}\right],
\,\frac{\displaystyle|z|}{\displaystyle2R_{CN}}\ll1\\[0.5cm]
\frac{\displaystyle4\pi^2}{\displaystyle|z|}\!\left[1\!-
\frac{\displaystyle1}{\displaystyle4}\!\left(\!\frac{\displaystyle2R_{CN}}{\displaystyle|z|}\!\right)^{\!\!2}\!
+\frac{\displaystyle9}{\displaystyle64}\!\left(\!\frac{\displaystyle2R_{CN}}{\displaystyle|z|}\!\right)^{\!\!4}\right]\!,
\,\frac{\displaystyle|z|}{\displaystyle2R_{CN}}\gg1\end{array}\right.\nonumber
\end{eqnarray}
and
\begin{eqnarray}
I_2\approx\left\{\begin{array}{l}\frac{\displaystyle4\pi}{\displaystyle\,\!R_{CN}}\!\left[\,
\frac{\displaystyle1}{\displaystyle2}\ln\!\left(\!\frac{\displaystyle4R_{CN}}{\displaystyle|z|}\right)\right.\\
\hskip2.cm\left.-1+\frac{\displaystyle3}{\displaystyle8}\left(\!\frac{\displaystyle|z|}{\displaystyle2R_{CN}}\right)^{\!2}\right],\;
\frac{\displaystyle|z|}{\displaystyle2R_{CN}}\ll1\\[0.5cm]
\frac{\displaystyle\pi^2}{\displaystyle4|z|}\left(\!\frac{\displaystyle2R_{CN}}{\displaystyle|z|}\!\right)^{\!\!2}\!\left[
1-\frac{\displaystyle3}{\displaystyle4}\left(\!\frac{\displaystyle2R_{CN}}{\displaystyle|z|}\!\right)^{\!\!2}\right],\;
\frac{\displaystyle|z|}{\displaystyle2R_{CN}}\gg1\end{array}\right.\nonumber
\end{eqnarray}
Plugging these $I_1$ and $I_2$ into Eq.~(\ref{Veffnew}) and
retaining only leading expansion terms yields
\begin{eqnarray}
V_{\mbox{\small{eff}}}(z)\approx\left\{\!\!\!\begin{array}{l}-\frac{\displaystyle\,\!e^2\left[1\!-\!\Delta_j(F)\right]}
{\displaystyle\pi\epsilon\!\,R_{CN}}\,\ln\!\left(\!\frac{\displaystyle4R_{CN}}{\displaystyle|z|}\!\right)\!,\;
\frac{\displaystyle|z|}{\displaystyle2R_{CN}}\ll1\\[0.5cm]
-\frac{\displaystyle e^2}{\displaystyle\epsilon|z|}\,,\;
\frac{\displaystyle|z|}{\displaystyle2R_{CN}}\gg1\end{array}\right.\label{Vefffin}
\end{eqnarray}

We see from Eq.~(\ref{Vefffin}) that, to the leading order in the
series expansion parameter, the perpendicular electrostatic field
does not affect the longitudinal electron-hole Coulomb potential
at large distances $|z|\gg2R_{CN}$, as one would expect. At short
distances $|z|\ll2R_{CN}$ the situation is different, however. The
potential decreases logarithmically with the field dependent
amplitude as $|z|$ goes down. The amplitude of the potential
decreases quadratically as the field increases [see
Eq.~(\ref{DeltaF})], thereby slowing down the potential fall-off
with decreasing $|z|$, or, in other words, making the potential
shallower as the field increases. Such a behavior can be uniformly
approximated for all $|z|$ by an appropriately chosen attractive
cusp-type cutoff potential with the field dependent cutoff
parameter. Indeed, consider the dimensionless function
$f(y)=-2R_{CN}\epsilon\,V_{\mbox{\small eff}}/e^2$ of the
dimensionless variable $y=|z|/2R_{CN}$. Then, according to
Eq.~(\ref{Vefffin}), one has
\[
f(y)=\left\{\begin{array}{l}\Phi_1(y)=\frac{\displaystyle2}{\displaystyle\pi}
\left[1-\Delta_j(F)\right]\ln\!\left(\frac{\displaystyle2}{\displaystyle y}\right),\;0<y\ll1\\[0.2cm]
\Phi_2(y)=\frac{\displaystyle1}{\displaystyle\,\!y}\,,\;y\gg1\end{array}\right.
\]
Now introduce the function
\begin{equation}
\Phi(y)=\frac{1}{y+y_0}\label{Phi}
\end{equation}
with the cutoff parameter $y_0$ selected in such a way as to
satisfy the condition $\Phi(1)=[\Phi_1(1)+\Phi_2(1)]/2$. This
yields
\begin{equation}
y_0=\frac{\pi-2\ln2\left[1-\Delta_j(F)\right]}{\pi+2\ln2\left[1-\Delta_j(F)\right]}\;.
\label{y0}
\end{equation}

\begin{figure}[t]
\epsfxsize=8.3cm\centering{\epsfbox{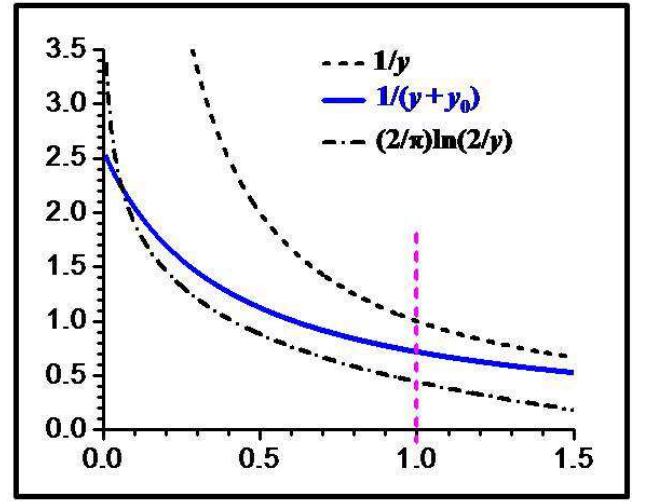}}\caption{(Color
online)~The dimensionless function (\ref{Phi}) with the zero-field
cutoff parameter (\ref{y0}). See text for details.}\label{fig8}
\end{figure}

Figure~\ref{fig8} shows the zero-field behavior of the $\Phi(y)$
function as compared to the corresponding $\Phi_1(y)$ and
$\Phi_2(y)$ functions. We see that $\Phi(y)$ gradually approaches
$\Phi_2(y)=1/y$ for increasing $y>1$. For decreasing $y<1$, on the
other hand, $\Phi(y)$ is very close to the logarithmic behavior as
given by $\Phi_1(y)$, with the exception that there is no
divergence at $y\sim0$ due to the presence of the cutoff. The
cutoff parameter (\ref{y0}) is field dependent, decreasing as the
field grows, which is consistent with the behavior of the original
potential (\ref{Vefffin}). Multiplying Eq.~(\ref{Phi}) by the
dimensional factor $-e^2/2R_{CN}\epsilon$ and putting
$y=|z|/2R_{CN}$, we obtain the attractive longitudinal cusp-type
cutoff potential (\ref{Vcutoff}) we build our analysis on in this
paper.


\begin{thebibliography}{99}
\bibitem{EndNote1}In real CNs, the existence of two equivalent energy
valleys in the 1st Brillouin zone, the $K$- and
$K^{\prime}$-valleys with opposite electron helicities about the
CN axis, results into dark and bright excitonic states in the
lowest energy spin-singlet manifold~\cite{AndoDarkExciton}. Since
the electric interaction does not involve spin variables, both
$K$- and $K^{\prime}$-valleys are affected equally by the
electrostatic field in our case, and the detailed structure of the
exciton wave function multiplet is not important. This is opposite
to the non-zero magnetostatic field case where the field affects
the $K$- and $K^{\prime}$-valleys differently either to brighten
the dark excitonic states~\cite{Srivastava}, or to create Landau
sublevels~\cite{Ando} for longitudinal and perpendicular
orientation, respectively.
\bibitem{EndNote2}We are based on the zero-exciton-temperature approximation
in here~\cite{Suna}, which is well justified because of the
exciton excitation energies much larger than $k_BT$ in~CNs. The
exciton Hamiltonian (\ref{Hex}) does not require the thermal
averaging over the exciton degrees of freedom then, yielding the
temperature independent total exciton energy (\ref{Ef}). One has
to keep in mind, however, that the exciton excitation energy can
be affected by the enviromental effect not under consideration in
here (see Ref.~\cite{Miyauchi}).
\end{thebibliography}
\end{document}